\documentclass[
  reprint,
  amsmath,amssymb,
  aps,
]{revtex4-2}

\usepackage{braket}
\usepackage{graphicx}
\usepackage{dcolumn}
\usepackage{bm}
\usepackage{subcaption}
\usepackage{multirow}

\begin{document}

\title{Symmetry classification of magnetic octupole current\\ based on multipole representation theory}

\author{Yuuga Takasu}
\author{Satoru Hayami}
\affiliation{Graduate School of Science, Hokkaido University, Sapporo 060-0810, Japan}

\date{\today}

\begin{abstract}
Magnetic octupole (MO) currents have recently attracted significant attention as a driving force for the N\'{e}el vector dynamics in \emph{d}-wave altermagnets, a new class of antiferromagnets that exhibit nonrelativistic spin-split band structures.
From a symmetry perspective, the MO includes an axial-dipole component analogous to that of the spin, making it essential to clarify how MO currents differ from spin currents.
We here investigate the correspondence between MO conductivities and electronic multipoles, which provide a unified and powerful framework for symmetry analysis.
We derive the multipole representation of the rank-five MO conductivity tensor and classify its symmetry-allowed components for all crystallographic point groups, in direct comparison with spin conductivity.
We show that time-reversal-even electric-type multipoles give rise to the dissipationless MO current, whereas time-reversal-odd magnetic-type multipoles generate dissipative MO current under an applied electric field.
Complementing this macroscopic analysis, the linear-response calculations for a microscopic tight-binding model demonstrate how MO conductivities are activated by symmetry lowering, exemplified by the symmetry reduction from $m\bar{3}m$ to $m\bar{3}$.
Our results elucidate the symmetry distinctions between MO currents and spin currents, and provide insights into their experimental identification.
\end{abstract}

\maketitle

\section{\label{sec:Introduction}Introduction}
In recent years, antiferromagnetic materials exhibiting nonrelativistic momentum-dependent spin-split band structures have been discovered~\cite{noda2016momentum,okugawa2018weakly,ahn2019antiferromagnetism,naka2019spin,hayami2019momentum,hayami2020bottom,yuan2020giant,naka2021perovskite,yuan2021prediction,yuan2021strong,Hayami_PhysRevB.105.024413}.
Such antiferromagnetic materials are referred to as \emph{altermagnets}~\cite{gonzalez2021efficient,vsmejkal2022giant,vsmejkal2022beyond}, and several intriguing phenomena free from spin--orbit coupling (SOC) have been proposed, such as the spin current generation~\cite{naka2019spin, hayami2022spin,sourounis2025efficient,ezawa2025third,naka2025altermagnetic} and linear and nonlinear piezo-magnetic effects~\cite{ma2021multifunctional,zhu2023multipiezo,aoyama2024piezomagnetic,mcclarty2024landau,wu2024valley,yershov2024fluctuation,chen2025strain,ogawa2025nonlinear,huyen2025anisotropic,naka2025nonrelativistic}.
In particular, \emph{d}-wave altermagnets, which exhibit nonrelativistic symmetric spin-split band structures characterized by the product of the second order of the wave vector $\bm{k}$ and collinear spin moment $\sigma$, $k_{i} k_{j} \sigma$ for $i,j=x,y,z$, have attracted considerable attention.
In these systems, phenomena arising from the magnetic octupole (MO) --- the lowest-rank magnetic anisotropy that couples to inhomogeneous magnetic fields~\cite{Bhowal_PhysRevX.14.011019, mcclarty2024landau, sato2025quantum, oike2025thermodynamic, schiff2025collinear} --- have been proposed, such as the nonlinear magnetoelectric effect~\cite{oike2024nonlinear}.

It has been revealed that MOs in \emph{d}-wave altermagnets can couple linearly to the N\'{e}el vector and act as secondary order parameters~\cite{mcclarty2024landau}.
By focusing on the similarities between this coupling and SOC, the concept of an MO torque induced by MO currents --- analogous to the spin-orbit torque induced by spin currents --- has been proposed~\cite{han2025harnessing}.
In parallel, theoretical studies have demonstrated N\'{e}el vector switching and domain wall dynamics in \emph{d}-wave altermagnets driven by such MO torques~\cite{han2025deterministic}.
These pioneering studies have brought MO currents into focus in the context of N\'{e}el-vector dynamics, and stimulated active investigation of their fundamental properties.
Furthermore, first-principles calculations have revealed finite MO Hall conductivities in a variety of 4\emph{d} and 5\emph{d} transition metals~\cite{baek2025magnetic}.

MO currents can be distinguished from spin currents in certain symmetry settings, even though both are described by axial response tensors~\cite{ko2025magnetic}.
This indicates that, despite sharing the same tensor character, the two currents are subject to different symmetry constraints on their components and can therefore appear differently in real materials.
Symmetry analyses of such response tensors are thus crucial not only for proposing experimental setups to identify MO currents --- by determining which components are allowed or forbidden under given crystal and magnetic point groups --- but also for elucidating the essential microscopic electronic degrees of freedom.

To systematically capture these symmetry constraints, we employ the framework of electronic multipoles~\cite{Santini_RevModPhys.81.807, kuramoto2009multipole, suzuki2018first, kusunose2022generalization, hayami2024unified}.
Multipoles can be categorized into four types based on their spatial inversion ($\mathcal{P}$) and time-reversal ($\mathcal{T}$) parities~\cite{dubovik1975multipole, kusunose2022generalization, hayami2024unified}: 
electric multipoles (E), corresponding to $\mathcal{T}$-even polar tensors; 
magnetic multipoles (M), corresponding to $\mathcal{T}$-odd axial tensors; 
magnetic toroidal multipoles (MT), corresponding to $\mathcal{T}$-odd polar tensors; 
and electric toroidal multipoles (ET), corresponding to $\mathcal{T}$-even axial tensors.
These multipoles are directly related to response tensors through symmetry considerations~\cite{hayami2018classification,watanabe2018group,yatsushiro2021multipole}.
The symmetries of the system restrict the activation of multipoles and, in turn, determine which components of the response tensors can remain finite under a given symmetry group. 
Moreover, symmetries link multipoles to microscopic electronic degrees of freedom~\cite{Kusunose_PhysRevB.107.195118}, so that the multipole framework serves as a powerful tool for exploring and designing unconventional response phenomena involving nontrivial couplings between electronic degrees of freedom.
For instance, unconventional longitudinal spin currents originating from an electric ferroaxial moment have been proposed in the context of an ET dipole, where the key microscopic ingredients have been identified~\cite{hayami2022electric}. 
For these reasons, multipole analysis is essential from both macroscopic and microscopic perspectives.

In the present study, we investigate the symmetry conditions of MO currents induced by electric fields.
We show the multipole representations of the MO conductivity tensors and clarify their symmetry distinctions from spin conductivity tensors under all crystallographic point groups (CPGs), thereby establishing key factors for experimental identification.
We also discuss the role of the time-reversal symmetry of the system on the MO conductivity.
We then perform linear response calculations for a tight-binding model to elucidate the microscopic properties of the MO conductivity associated with the symmetry lowering from the point group $m\bar{3}m$ to $m\bar{3}$.

This paper is organized as follows.
In Sec.~\ref{sec:Symmetry of MO conductivity}, we show the relationships between the MO conductivity tensors and the multipoles.
We compare the obtained results with those of the spin conductivity tensors (i) for the case where the multipoles activated expressed by cubic harmonics (\emph{c}-multipoles), which are relevant to the CPG $m\bar{3}m$ and its subgroups, and (ii) for all CPGs.
Then, we present the role of $\mathcal{T}$ symmetry on the MO conductivity and show the relevance to multipoles from the viewpoint of the dissipation based on the Kubo formula.
In Sec.~\ref{sec:Model Analysis},we construct a basic tight-binding model and present numerical results to demonstrate symmetry aspects arising from the microscopic details of the system.
Section~\ref{sec:Summary} provides a summary of the paper.
In Appendix~\ref{app:Angle dependence of multipole}, we present the angle dependences of \emph{c}-multipoles and multipoles expressed by tesseral harmonics (\emph{t}-multipoles) relevant to the hexagonal point group $6/mmm$ and its subgroups.
Appendix~\ref{app:Multipole expressions of MO conductivities} presents the relationship between the tensor components of the MO conductivities and relevant multipoles.
Appendix~\ref{app:Comparison between the magnetic octupole conductivities and spin conductivities for t-multipoles} summarizes the same comparison as in Sec.~\ref{sec:Comparison between the magnetic octupole and spin conductivities} for the \emph{t}-multipoles.
Finally, Appendix~\ref{app:Hopping Hamiltonian} lists the explicit forms of the hopping matrices used in the numerical calculations.

\section{\label{sec:Symmetry of MO conductivity}Symmetry of Magnetic Octupole Conductivity}
In this section, we investigate the symmetry of the MO conductivity $\sigma_{i; j}^{mn \alpha}$, which is characterized as a rank-5 axial tensor as follows:
\begin{equation}
  J_{i}^{mn \alpha} = \sigma_{i; j}^{mn \alpha} E_{j},
  \label{eq:MO conductivity}
\end{equation}
where $J_{i}^{mn \alpha}$ and $E_{j}$ for $i, j, m, n, \alpha = x, y, z$ denote the MO current and the electric field, respectively.
Here, $J_{i}^{mn \alpha}$ is defined by using the MO operator $O_{mn}^{\alpha} = \{L_{m}, L_{n}\}_{+} S_{\alpha} / (2\hbar^{2})$ as $J_{i}^{mn \alpha} \equiv \{O_{mn}^{\alpha}, v_{i}\}_{+} / 2$~\cite{han2025harnessing,baek2025magnetic,ko2025magnetic}, which is analous to the spin current~\cite{sinova2004universal}. 
The symbols $\{\cdot,\cdot\}_{+}$, $v_{i}$, $L_{m}$, $S_{\alpha}$, and $\hbar$ denote the anticommutator, velocity, orbital angular momentum, spin angular momentum, and the reduced Planck constant, respectively. 

In the following, we analyze the symmetry of the MO conductivities using multipole representations.
First, we discuss the correspondence between the MO conductivities and the multipoles by decomposing the MO conductivities into the Ohmic and Hall parts in Sec.~\ref{sec:Correspondence to multipole}.
We then apply this method to the spin conductivities and compare the results with those of the MO conductivities in Sec.~\ref{sec:Comparison between the magnetic octupole and spin conductivities}.
Moreover, we demonstrate the distinctions among them under the 32 CPGs in Sec.~\ref{sec:Classification under 32 point groups}.
In Sec.~\ref{sec:Time-reversal property}, we show that multipoles with different parities under $\mathcal{T}$ symmetry give rise to different contributions with respect to the dissipation based on the Kubo formula.

\subsection{\label{sec:Correspondence to multipole}Correspondence to Multipole}
We first show the correspondence between the MO conductivity tensors $\sigma_{i; j}^{mn \alpha}$ and multipoles.
We introduce the symbols $X$ and $Y$ to distinguish polar and axial tensors, respectively.
We assume that the time-reversal symmetry is present unless stated otherwise; see Sec.~\ref{sec:Time-reversal property} for the role of time-reversal symmetry on the MO conductivity.
Depending on the crystal system, we use two types of multipole notation: One is the \emph{c}-multipole for the cubic point group and its subgroups, and the other is the \emph{t}-multipole for the hexagonal point group and its subgroups.
Following Refs.~\cite{hayami2018classification,yatsushiro2021multipole}, the notations for \emph{c}- (or \emph{t}-) multipole are given by 
$X_{0}$ for the rank-0 monopole, $\bm{X} = \{X_{x}, X_{y}, X_{z}\}$ for the rank-1 dipole, $X_{ij} = \{X_{u}, X_{v}, X_{yz}, X_{zx}, X_{xy}\}$ for the rank-2 quadrupole,
$X_{ijk} = \{X_{xyz}, X^{\alpha}_{x}, X^{\alpha}_{y}, X^{\alpha}_{z}, X^{\beta}_{x}, X^{\beta}_{y}, X^{\beta}_{z}\}$ ($= \{X_{xyz}, X_{z}^{\alpha}, X_{z}^{\beta}, X_{3u}, X_{3v}, X_{3a}, X_{3b}\}$) for the rank-3 octupole,
$X_{ijkl} = \{X_{4}, X_{4u}, X_{4v}, X^{\alpha}_{4x}, X^{\alpha}_{4y}, X^{\alpha}_{4z}, X^{\beta}_{4x}, X^{\beta}_{4y}, X^{\beta}_{4z}\}$ ($= \{X_{40}, X_{4a}, X_{4b}, X_{4u}^{\alpha}, X_{4v}^{\alpha}, X_{4u}^{\beta 1}, X_{4v}^{\beta 1}, X_{4u}^{\beta 2}, X_{4v}^{\beta 2}\}$) for the rank-4 hexadecapole,
and $X_{ijklm} = \{X_{5u}, X_{5v}, X^{\alpha1}_{5x}, X^{\alpha1}_{5y}, X^{\alpha1}_{5z}, X^{\alpha2}_{5x}, X^{\alpha2}_{5y}, X^{\alpha2}_{5z}, X^{\beta}_{5x}, X^{\beta}_{5y}, X^{\beta}_{5z}\}$ ($= \{X_{50}, X_{5a}, X_{5b}, X_{5u}^{\alpha 1}, X_{5v}^{\alpha 1}, X_{5u}^{\alpha 2}, X_{5v}^{\alpha 2}, X_{5u}^{\beta 1}, X_{5v}^{\beta 1}, X_{5u}^{\beta 2}, X_{5v}^{\beta 2}\}$) for the rank-5 dotriacontapole.
It is noted that \emph{t}-multipoles from rank 0 to rank 2 are the same as \emph{c}-multipoles.
Angular dependences of multipoles are shown in Table~\ref{tab:angle dependences of multipoles} in Appendix~\ref{app:Angle dependence of multipole}.

Starting from the decomposition of the orbital part in $O_{mn}^{\alpha}$, the multipole correspondence is given by
\begin{equation}
  \{L_{m}, L_{n}\}_{+} \leftrightarrow X_{0} \oplus X_{ij},
  \label{eq:multipole decomposition of orbital part}
\end{equation}
where $X_{0}$ and $X_{ij}$ corresponds to $\bm{L} \cdot \bm{L}$ and $L_{m}L_{n} + L_{n}L_{m}$, respectively.
Thus, the quantity $\{L_{m}, L_{n}\}_{+}$ includes the rank-0 and rank-2 multipoles; it is noted that no rank-1 multipoles appear owing to the symmetric tensor property of $\{L_{m}, L_{n}\}_{+}$.
Then, since the spin corresponds to a rank-1 axial tensor, the multipole representation of $O_{mn}^{\alpha}$ is given by
\begin{equation}
  O_{mn}^{\alpha} \leftrightarrow (X_{0} \oplus X_{ij}) \otimes \bm{Y} = \bm{Y} \oplus \bm{Y}' \oplus X_{ij} \oplus Y_{ijk},
  \label{eq:multipole decomposition of MO}
\end{equation}
where $\bm{Y}$ and $\bm{Y}'$ originate from $X_{0} \otimes \bm{Y}$ and $X_{ij} \otimes \bm{Y}$, respectively, the latter of which is referred to as the anisotropic dipole~\cite{yamasaki2020augmented, Hayami_PhysRevB.103.L180407, Sasabe_PhysRevLett.131.216501}.
Thus, the rank-3 tensor generally includes two types of dipoles, one quadrupole, and one octupole~\cite{urru:annphys2022}.

In a similar manner, since both the current $J_i$ (or velocity $v_i$) and the electric field $E_j$ transform as polar dipoles, their product can be decomposed as 
\begin{equation}
  J_i E_j \leftrightarrow X_{0} \oplus \bm{Y} \oplus X_{ij},
  \label{eq:multipole decomposition of currents}
\end{equation}
where the rank-0 $X_0$ and rank-2 $X_{ij}$ terms constitute the symmetric components ($J_i E_j + J_j  E_i$), whereas the rank-1 $\bm{Y}$ represents the antisymmetric component ($J_i E_j - J_j  E_i$).
In the context of electric conductivity, the former is called the Ohmic contribution, while the latter is called the Hall contribution.
By combining it with the MO in Eq.~(\ref{eq:multipole decomposition of MO}), the correspondence between rank-5 MO conductivity tensor and multipoles are obtained.
The relevant multipoles for the Ohmic contribution of the MO conductivity are given by 
\begin{equation}
  (J_i E_j + J_j  E_i) O_{mn}^{\alpha} \leftrightarrow X_{0} \oplus 6 \bm{Y} \oplus 5 X_{ij} \oplus 5 Y_{ijk} \oplus 2 X_{ijkl} \oplus Y_{ijklm}, 
  \label{eq:multipole decomposition of MO Ohmic conductivity}
\end{equation}
while those for Hall contribution of the MO conductivity are given by 
\begin{equation}
  (J_i E_j - J_j  E_i) O_{mn}^{\alpha} \leftrightarrow 2 X_{0} \oplus 3 \bm{Y} \oplus 4 X_{ij} \oplus 2 Y_{ijk} \oplus X_{ijkl}, 
  \label{eq:multipole decomposition of MO Hall conductivity}
\end{equation}
where coefficients of multipoles stand for independent multipole numbers.
Thus, the MO conductivity tensor $\sigma_{i; j}^{mn \alpha}$ contains polar even-rank and axial odd-rank multipoles up to rank 5. 
From the spatial inversion parity, $X$ corresponds to E (MT) multipoles ($X=Q$ or $T$), while $Y$ corresponds to ET (M) multipoles ($Y=G$ or $M$).

Next, we express the tensor components in terms of multipoles.
As described above, $\sigma_{i; j}^{mn \alpha}$ is divided into the Ohmic part $\sigma_{ij}^{{mn\alpha} (\mathrm{O})}$ and the Hall part $\sigma_{ij}^{mn\alpha (\mathrm{H})}$ as follows: 
\begin{align}
  \sigma_{i; j}^{mn\alpha} &= \sigma_{ij}^{mn\alpha (\mathrm{O})} + \sigma_{ij}^{mn\alpha (\mathrm{H}),} \\
  \sigma_{ij}^{mn\alpha (\mathrm{O})} &\equiv  \tfrac{1}{2} (\sigma_{i; j}^{mn\alpha} + \sigma_{j; i}^{mn\alpha}), \\
  \sigma_{ij}^{mn\alpha (\mathrm{H})} &\equiv \tfrac{1}{2} (\sigma_{i; j}^{mn\alpha} - \sigma_{j; i}^{mn\alpha}),
\end{align}
where the Ohmic and Hall parts are characterized by the symmetric and antisymmetric tensors for the interchange of $i$ and $j$, respectively, i.e., $\sigma_{ij}^{{mn\alpha} (\mathrm{O})}=\sigma_{ji}^{{mn\alpha} (\mathrm{O})}$ and $\sigma_{ij}^{{mn\alpha} (\mathrm{H})}=-\sigma_{ji}^{{mn\alpha} (\mathrm{H})}$.
The Ohmic conductivity tensor is represented by the $18 \times 6$ matrix as 
\begin{equation}
  \left(\begin{array}{cccccc}
    \sigma_{xx}^{xxx (\mathrm{H})} & \sigma_{yy}^{xxx (\mathrm{H})} & \sigma_{zz}^{xxx (\mathrm{H})} & \sigma_{yz}^{xxx (\mathrm{H})} & \sigma_{zx}^{xxx (\mathrm{H})} & \sigma_{xy}^{xxx (\mathrm{H})} \\
    \sigma_{xx}^{yyx (\mathrm{H})} & \sigma_{yy}^{yyx (\mathrm{H})} & \sigma_{zz}^{yyx (\mathrm{H})} & \sigma_{yz}^{yyx (\mathrm{H})} & \sigma_{zx}^{yyx (\mathrm{H})} & \sigma_{xy}^{yyx (\mathrm{H})} \\
    \sigma_{xx}^{zzx (\mathrm{H})} & \sigma_{yy}^{zzx (\mathrm{H})} & \sigma_{zz}^{zzx (\mathrm{H})} & \sigma_{yz}^{zzx (\mathrm{H})} & \sigma_{zx}^{zzx (\mathrm{H})} & \sigma_{xy}^{zzx (\mathrm{H})} \\
    \sigma_{xx}^{yzx (\mathrm{H})} & \sigma_{yy}^{yzx (\mathrm{H})} & \sigma_{zz}^{yzx (\mathrm{H})} & \sigma_{yz}^{yzx (\mathrm{H})} & \sigma_{zx}^{yzx (\mathrm{H})} & \sigma_{xy}^{yzx (\mathrm{H})} \\
    \sigma_{xx}^{zxx (\mathrm{H})} & \sigma_{yy}^{zxx (\mathrm{H})} & \sigma_{zz}^{zxx (\mathrm{H})} & \sigma_{yz}^{zxx (\mathrm{H})} & \sigma_{zx}^{zxx (\mathrm{H})} & \sigma_{xy}^{zxx (\mathrm{H})} \\
    \sigma_{xx}^{xyx (\mathrm{H})} & \sigma_{yy}^{xyx (\mathrm{H})} & \sigma_{zz}^{xyx (\mathrm{H})} & \sigma_{yz}^{xyx (\mathrm{H})} & \sigma_{zx}^{xyx (\mathrm{H})} & \sigma_{xy}^{xyx (\mathrm{H})} \\
    \sigma_{xx}^{xxy (\mathrm{H})} & \sigma_{yy}^{xxy (\mathrm{H})} & \sigma_{zz}^{xxy (\mathrm{H})} & \sigma_{yz}^{xxy (\mathrm{H})} & \sigma_{zx}^{xxy (\mathrm{H})} & \sigma_{xy}^{xxy (\mathrm{H})} \\
    \sigma_{xx}^{yyy (\mathrm{H})} & \sigma_{yy}^{yyy (\mathrm{H})} & \sigma_{zz}^{yyy (\mathrm{H})} & \sigma_{yz}^{yyy (\mathrm{H})} & \sigma_{zx}^{yyy (\mathrm{H})} & \sigma_{xy}^{yyy (\mathrm{H})} \\
    \sigma_{xx}^{zzy (\mathrm{H})} & \sigma_{yy}^{zzy (\mathrm{H})} & \sigma_{zz}^{zzy (\mathrm{H})} & \sigma_{yz}^{zzy (\mathrm{H})} & \sigma_{zx}^{zzy (\mathrm{H})} & \sigma_{xy}^{zzy (\mathrm{H})} \\
    \sigma_{xx}^{yzy (\mathrm{H})} & \sigma_{yy}^{yzy (\mathrm{H})} & \sigma_{zz}^{yzy (\mathrm{H})} & \sigma_{yz}^{yzy (\mathrm{H})} & \sigma_{zx}^{yzy (\mathrm{H})} & \sigma_{xy}^{yzy (\mathrm{H})} \\
    \sigma_{xx}^{zxy (\mathrm{H})} & \sigma_{yy}^{zxy (\mathrm{H})} & \sigma_{zz}^{zxy (\mathrm{H})} & \sigma_{yz}^{zxy (\mathrm{H})} & \sigma_{zx}^{zxy (\mathrm{H})} & \sigma_{xy}^{zxy (\mathrm{H})} \\
    \sigma_{xx}^{xyy (\mathrm{H})} & \sigma_{yy}^{xyy (\mathrm{H})} & \sigma_{zz}^{xyy (\mathrm{H})} & \sigma_{yz}^{xyy (\mathrm{H})} & \sigma_{zx}^{xyy (\mathrm{H})} & \sigma_{xy}^{xyy (\mathrm{H})} \\
    \sigma_{xx}^{xxz (\mathrm{H})} & \sigma_{yy}^{xxz (\mathrm{H})} & \sigma_{zz}^{xxz (\mathrm{H})} & \sigma_{yz}^{xxz (\mathrm{H})} & \sigma_{zx}^{xxz (\mathrm{H})} & \sigma_{xy}^{xxz (\mathrm{H})} \\
    \sigma_{xx}^{yyz (\mathrm{H})} & \sigma_{yy}^{yyz (\mathrm{H})} & \sigma_{zz}^{yyz (\mathrm{H})} & \sigma_{yz}^{yyz (\mathrm{H})} & \sigma_{zx}^{yyz (\mathrm{H})} & \sigma_{xy}^{yyz (\mathrm{H})} \\
    \sigma_{xx}^{zzz (\mathrm{H})} & \sigma_{yy}^{zzz (\mathrm{H})} & \sigma_{zz}^{zzz (\mathrm{H})} & \sigma_{yz}^{zzz (\mathrm{H})} & \sigma_{zx}^{zzz (\mathrm{H})} & \sigma_{xy}^{zzz (\mathrm{H})} \\
    \sigma_{xx}^{yzz (\mathrm{H})} & \sigma_{yy}^{yzz (\mathrm{H})} & \sigma_{zz}^{yzz (\mathrm{H})} & \sigma_{yz}^{yzz (\mathrm{H})} & \sigma_{zx}^{yzz (\mathrm{H})} & \sigma_{xy}^{yzz (\mathrm{H})} \\
    \sigma_{xx}^{zxz (\mathrm{H})} & \sigma_{yy}^{zxz (\mathrm{H})} & \sigma_{zz}^{zxz (\mathrm{H})} & \sigma_{yz}^{zxz (\mathrm{H})} & \sigma_{zx}^{zxz (\mathrm{H})} & \sigma_{xy}^{zxz (\mathrm{H})} \\
    \sigma_{xx}^{xyz (\mathrm{H})} & \sigma_{yy}^{xyz (\mathrm{H})} & \sigma_{zz}^{xyz (\mathrm{H})} & \sigma_{yz}^{xyz (\mathrm{H})} & \sigma_{zx}^{xyz (\mathrm{H})} & \sigma_{xy}^{xyz (\mathrm{H})} \\
  \end{array}\right),
  \label{eq:index table of MO Ohmic conductivities}
\end{equation}
where the column and row stand for the indices of the MO and $J_i E_j$, respectively.
The Hall conductivity tensor is represented by the $18 \times 3$ matrix, which is given by
\begin{equation}
  \left(\begin{array}{ccc}
    \sigma_{yz}^{xxx (\mathrm{H})} & \sigma_{zx}^{xxx (\mathrm{H})} & \sigma_{xy}^{xxx (\mathrm{H})} \\
    \sigma_{yz}^{yyx (\mathrm{H})} & \sigma_{zx}^{yyx (\mathrm{H})} & \sigma_{xy}^{yyx (\mathrm{H})} \\
    \sigma_{yz}^{zzx (\mathrm{H})} & \sigma_{zx}^{zzx (\mathrm{H})} & \sigma_{xy}^{zzx (\mathrm{H})} \\
    \sigma_{yz}^{yzx (\mathrm{H})} & \sigma_{zx}^{yzx (\mathrm{H})} & \sigma_{xy}^{yzx (\mathrm{H})} \\
    \sigma_{yz}^{zxx (\mathrm{H})} & \sigma_{zx}^{zxx (\mathrm{H})} & \sigma_{xy}^{zxx (\mathrm{H})} \\
    \sigma_{yz}^{xyx (\mathrm{H})} & \sigma_{zx}^{xyx (\mathrm{H})} & \sigma_{xy}^{xyx (\mathrm{H})} \\
    \sigma_{yz}^{xxy (\mathrm{H})} & \sigma_{zx}^{xxy (\mathrm{H})} & \sigma_{xy}^{xxy (\mathrm{H})} \\
    \sigma_{yz}^{yyy (\mathrm{H})} & \sigma_{zx}^{yyy (\mathrm{H})} & \sigma_{xy}^{yyy (\mathrm{H})} \\
    \sigma_{yz}^{zzy (\mathrm{H})} & \sigma_{zx}^{zzy (\mathrm{H})} & \sigma_{xy}^{zzy (\mathrm{H})} \\
    \sigma_{yz}^{yzy (\mathrm{H})} & \sigma_{zx}^{yzy (\mathrm{H})} & \sigma_{xy}^{yzy (\mathrm{H})} \\
    \sigma_{yz}^{zxy (\mathrm{H})} & \sigma_{zx}^{zxy (\mathrm{H})} & \sigma_{xy}^{zxy (\mathrm{H})} \\
    \sigma_{yz}^{xyy (\mathrm{H})} & \sigma_{zx}^{xyy (\mathrm{H})} & \sigma_{xy}^{xyy (\mathrm{H})} \\
    \sigma_{yz}^{xxz (\mathrm{H})} & \sigma_{zx}^{xxz (\mathrm{H})} & \sigma_{xy}^{xxz (\mathrm{H})} \\
    \sigma_{yz}^{yyz (\mathrm{H})} & \sigma_{zx}^{yyz (\mathrm{H})} & \sigma_{xy}^{yyz (\mathrm{H})} \\
    \sigma_{yz}^{zzz (\mathrm{H})} & \sigma_{zx}^{zzz (\mathrm{H})} & \sigma_{xy}^{zzz (\mathrm{H})} \\
    \sigma_{yz}^{yzz (\mathrm{H})} & \sigma_{zx}^{yzz (\mathrm{H})} & \sigma_{xy}^{yzz (\mathrm{H})} \\
    \sigma_{yz}^{zxz (\mathrm{H})} & \sigma_{zx}^{zxz (\mathrm{H})} & \sigma_{xy}^{zxz (\mathrm{H})} \\
    \sigma_{yz}^{xyz (\mathrm{H})} & \sigma_{zx}^{xyz (\mathrm{H})} & \sigma_{xy}^{xyz (\mathrm{H})} \\
  \end{array}\right),
  \label{eq:index table of MOHE}
\end{equation}
where the column and row stand for the indices of the MO and $(\bm{E}\times \bm{J})$, respectively.

The correspondence between the MO conductivity and the multipoles is obtained through the tensor-product relation between multipoles:
\begin{equation}
  \hat{X}_{lq}
  = \sum_{q_{1}, q_{2}}
  \braket{l_{1}, l_{2}; q_{1}, q_{2} | l_{1}, l_{2}; l, q}
  \hat{X}_{l_{1} q_{1}} \hat{X}_{l_{2} q_{2}},
  \label{eq:coupling of multipoles}
\end{equation}
where $l$ stands for the angular momentum quantum number, $q$ represents the magnetic angular momentum quantum number, and $\braket{l_{1}, l_{2}; q_{1}, q_{2} | l_{1}, l_{2}; l, q}$ denotes the Clebsch-Gordan (CG) coefficient.  

Based on this expression, we sequentially apply tensor products to the quantities associated with the MO conductivity tensor.
First, the coupling between $L_{m}$ and $L_{n}$ leads to the polar monopole and quadrupole $\hat{X}_{l' q'}$ for $l' = 0, 2$ shown in Eq.~(\ref{eq:multipole decomposition of orbital part}):
\begin{align}
  \nonumber
  \{L_{m}, L_{n}\}_{+} \leftrightarrow & \hat{X}_{l'q'} \\
  = & \sum_{q_{m}, q_{n}} \sum_{m, n} \braket{1, 1; q_{m}, q_{n} | 1, 1; l', q'} C_{m} C_{n} Y_{m} Y_{n},
  \label{eq:multipole decomposition of orbital part with CG}
\end{align}
where we take the linear transformation between the dipole $X_{a}$ for $a = x, y, z$ in the real-number expressoin and $\hat{X}_{1 q_{a}}$ in the complex-number expression, which is connected by $\hat{X}_{1 q_{a}} = \sum_{a} C_{a} X_{a}$.
Then, the multipole decomposition of the MO term $\hat{X}_{l'' q''}$ for $l'' = 1, 2, 3$ is derived by combining Eq.~(\ref{eq:multipole decomposition of orbital part with CG}) with the axial dipole $\hat{Y}_{1 q_{\alpha}}$ as
\begin{align}
  \nonumber
  O_{mn}^{\alpha} & \leftrightarrow \hat{X}_{l'' q''} \\
  \nonumber
  & = \sum_{q', q_{\alpha}} \braket{l', 1; q', q_{\alpha} | l', 1; l'', q''} \hat{X}_{l' q'} \hat{Y}_{1 q_{\alpha}}\\
  \nonumber
  & = \sum_{q', q_{\alpha}} \sum_{q_{m}, q_{n}} \sum_{m, n, \alpha} \braket{l', 1; q', q_{\alpha} | l', 1; l'', q''} \\
  \nonumber
  & \times \braket{1, 1; q_{m}, q_{n} | 1, 1; l', q'} \\
  & \times C_{m} C_{n} C_{\alpha} Y_{m} Y_{n} Y_{\alpha}.
\end{align}
Similarly, the combination of $J_{i}$ and $E_{j}$ yields the multipoles $\hat{X}_{l''' q'''}$ for $l''' = 0, 1, 2$:
\begin{align}
  \nonumber
  J_{i} E_{j} & \leftrightarrow \hat{X}_{l''' q'''} \\
  & = \sum_{q_{i}, q_{j}} \sum_{i, j} \braket{1, 1; q_{i}, q_{j} | 1, 1; l''', q'''} C_{i} C_{j} X_{i} X_{j},
\end{align}
where the monopoles and quadrupoles (or dipoles) correspond to the MO Ohmic (or Hall) conductivities as mentioned above.
Thus, the multipole component $\hat{X}_{l,q}$ of the MO Ohmic (Hall) conductivity is obtained by combining $\hat{X}_{l'' q''}$ and $\hat{X}_{l''' q'''}$ for $l''' = 0, 2$ ($l''' = 1$) in the form:
\begin{align}
  \nonumber
  \sigma_{ij}^{mn\alpha(\mathrm{O/H})} \leftrightarrow & \hat{X}_{l,q} \\
  \nonumber
  = & \sum_{q'', q'''} \braket{l'', l'''; q'', q''' | l'', l'''; l, q} \hat{X}_{l'' q''} \hat{X}_{l''' q'''} \\
  \nonumber
  = & \sum_{q'', q'''} \sum_{q_{i}, q_{j}} \sum_{q', q_{\alpha}} \sum_{q_{m}, q_{n}} \\
  \nonumber
  & \braket{l'', l'''; q'', q''' | l'', l'''; l, q} \\
  \nonumber
  \times & \braket{1, 1; q_{i}, q_{j} | 1, 1; l''', q'''} \\
  \nonumber
  \times & \braket{l', 1; q', q_{\alpha} | l', 1; l'', q''} \\
  \nonumber
  \times & \braket{1, 1; q_{m}, q_{n} | 1, 1; l', q'} \\
  \times & C_{i} C_{j} C_{m} C_{n} C_{\alpha} X_{i} X_{j} Y_{m} Y_{n} Y_{\alpha}.
  \label{eq:multipole decomposition of MO conductivities with CG}
\end{align}
In Appendix \ref{app:Multipole expressions of MO conductivities}, we present the MO conductivities for all \emph{c}- and \emph{t}-multipoles in the forms of Eqs.~(\ref{eq:index table of MO Ohmic conductivities}) and (\ref{eq:index table of MOHE}), given as the coefficients of $X_{i} X_{j} Y_{m} Y_{n} Y_{\alpha}$ for each multipole, derived as linear combinations of Eq.~(\ref{eq:multipole decomposition of MO conductivities with CG}).

\subsection{\label{sec:Comparison between the magnetic octupole and spin conductivities}Comparison between the magnetic octupole and spin conductivities}

We show the differences between $\sigma_{i; j}^{mn\alpha}$ and the spin conductivity $\sigma_{i; j}^{\alpha}$ from a symmetry perspective.
The spin conductivity appears in the linear response regime as
\begin{equation}
  J_{j}^{\alpha} = \sigma_{i; j}^{\alpha} E_{j},
\end{equation}
where $J_{j}^{\alpha}$ is defined as $\{S_{\alpha}, v_{i}\}_{+}/2$~\cite{sinova2004universal}.
Since the spin transforms as axial dipole, the multipole correspondence of the spin Ohmic (or Hall) conductivities are obtained by combinations of axial dipole $\bm{Y}$ and $X_{0}, X_{ij}$ (or $\bm{Y}$) components in Eq.~(\ref{eq:multipole decomposition of currents})~\cite{hayami2018classification,yatsushiro2021multipole}.
The Ohmic part of the spin conductivities is given by rank-1 to rank-3 multipoles:
\begin{equation}
  \bm{Y} \otimes (X_{0} \oplus X_{ij}) = 2 \bm{Y} \oplus X_{ij} \oplus Y_{ijk},
  \label{eq:multipole decomposition of spin Ohmic conductivity}
\end{equation}
while the Hall part is given by rank-0 to rank-2 multipoles:
\begin{equation}
  \bm{Y} \otimes \bm{Y} = X_{0} \oplus \bm{Y} \oplus X_{ij}.
  \label{eq:multipole decomposition of spin Hall conductivity}
\end{equation}

We compare the nonvanishing tensor components of the MO and spin conductivities from the viewpoint of active \emph{c}-multipole.
To clarify the distinction between the two, we show which tensor components of the MO conductivity and the spin conductivity become finite under the corresponding active multipoles ($X$ or $Y$), as summarized in the case of Ohmic contributions in Tables~\ref{tab:MO spin Ohmic} and \ref{tab:MO Ohmic} and Hall contributions in Tables~\ref{tab:MO spin Hall} and \ref{tab:MO Hall}.
In the tables, we introduce the following matrix representation of nonzero MO conductivities:
\begin{equation}
  \scalebox{1}{$
  \left[\begin{array}{c|c|c}
    \{O_{mn}^{\alpha}\}_{xx} & \{O_{mn}^{\alpha}\}_{xy} & \{O_{mn}^{\alpha}\}_{xz} \\ \hline
    \{O_{mn}^{\alpha}\}_{yx} & \{O_{mn}^{\alpha}\}_{yy} & \{O_{mn}^{\alpha}\}_{yz} \\ \hline
    \{O_{mn}^{\alpha}\}_{zx} & \{O_{mn}^{\alpha}\}_{zy} & \{O_{mn}^{\alpha}\}_{zz}
  \end{array}\right],
  $}
  \label{eq:nonzero MO conductivity}
\end{equation}
where $\{O_{mn}^{\alpha}\}_{ij}$ lists nonzero tensor components of MO Ohmic/Hall conductivities $ \sigma_{ij}^{mn\alpha(\mathrm{O/H})}$ for the input $i$-directional electric field and the output $j$-directoinal current.
Similarly, the nonzero components of the spin conductivities are summarized as
\begin{equation}
  \scalebox{1}{$
  \left[\begin{array}{c|c|c}
    \{S_{\alpha}\}_{xx} & \{S_{\alpha}\}_{xy} & \{S_{\alpha}\}_{xz} \\ \hline
    \{S_{\alpha}\}_{yx} & \{S_{\alpha}\}_{yy} & \{S_{\alpha}\}_{yz} \\ \hline
    \{S_{\alpha}\}_{zx} & \{S_{\alpha}\}_{zy} & \{S_{\alpha}\}_{zz}
  \end{array}\right],
  $}
  \label{eq:nonzero spin conductivity}
\end{equation}
where $\{S_{\alpha}\}_{ij}$ list nonzero tensor components of spin Ohmic/Hall conductivities $\sigma_{ij}^{\alpha(\mathrm{O/H})}$.
The corresponding tables for the \emph{t}-multipoles, i.e., Tables~\ref{tab:MO spin Ohmic tesseral}, \ref{tab:MO Ohmic tesseral}, and \ref{tab:MO Hall tesseral}, are provided in Appendix~\ref{app:Comparison between the magnetic octupole conductivities and spin conductivities for t-multipoles}.

\begin{widetext}
  \begin{table*}
  \caption{\raggedright
    Symmetry-allowed MO Ohmic conductivity tensors and spin Ohmic conductivity tensors with activations from rank-1 to -3 c-multipoles (\emph{c}-MP).
    The matrix elements only the upper-triangular parts are shown, since the lower-triangular components are given by satisfying
    $\sigma_{ij}^{mn\alpha(\mathrm{O})} = \sigma_{ji}^{mn\alpha(\mathrm{O})}$, 
    $\sigma_{ij}^{mn\alpha(\mathrm{H})} = -\sigma_{ji}^{mn\alpha(\mathrm{H})}$, 
    $\sigma_{ij}^{\alpha(\mathrm{O})} = \sigma_{ji}^{\alpha(\mathrm{O})}$, and
    $\sigma_{ij}^{\alpha(\mathrm{H})} = -\sigma_{ji}^{\alpha(\mathrm{H})}$.
    The symbol $-$ indicates no symmetry-allowed components.
    The abbreviation as $\bm{O}_{mn}^{\alpha} = \{O_{yz}^{x}, O_{zx}^{y}, O_{xy}^{z}\}$, $\bm{O}_{yz} = \{O_{yz}^{x}, O_{yz}^{y}, O_{yz}^{z}\}$, $\bm{O}_{zx} = \{O_{zx}^{x}, O_{zx}^{y}, O_{zx}^{z}\}$, $\bm{O}_{xy} = \{O_{xy}^{x}, O_{xy}^{y}, O_{xy}^{z}\}$, $\bm{O}^{i} = \{O_{xx}^{i}, O_{yy}^{i}, O_{zz}^{i}\}$ for $i = x, y, z$, $\mathrm{all} = \{\bm{O}_{yz}, \bm{O}_{zx}, \bm{O}_{xy}, \bm{O}^{x}, \bm{O}^{y}, \bm{O}^{z}\}$ is used for notational simplicity.
  }
  \centering
  \begin{tabular}{c|c|c}
    \emph{c}-MP & MO conductivity & spin conductivity \\ \hline \hline
    $Y_{x}$ & \scalebox{1}{$\begin{bmatrix}\begin{array}{c|c|c} O_{xy}^{y}, O_{zx}^{z}, \bm{O}^{x} & O_{xy}^{x}, O_{yz}^{z}, \bm{O}^{y} & O_{zx}^{x}, O_{yz}^{y}, \bm{O}^{z} \\ \hline & O_{xy}^{y}, O_{zx}^{z}, \bm{O}^{x} & \bm{O}_{lm}^{\alpha} \\ \hline & & O_{xy}^{y}, O_{zx}^{z}, \bm{O}^{x}\end{array}\end{bmatrix}$} & \scalebox{1}{$\left[\begin{array}{c|c|c} S_{x} & S_{y} & S_{z} \\ \hline & S_{x} & - \\ \hline & & S_{x} \\ \end{array}\right]$}\rule[-18pt]{0pt}{40pt} \\ \hline
    $Y_{y}$ & \scalebox{1}{$\begin{bmatrix}\begin{array}{c|c|c} O_{xy}^{x}, O_{yz}^{z}, \bm{O}^{y} & O_{xy}^{y}, O_{zx}^{z}, \bm{O}^{x} & \bm{O}_{lm}^{\alpha} \\ \hline & O_{xy}^{x}, O_{yz}^{z}, \bm{O}^{y} & O_{zx}^{x}, O_{yz}^{y}, \bm{O}^{z} \\ \hline & & O_{xy}^{x}, O_{yz}^{z}, \bm{O}^{y}\end{array}\end{bmatrix}$} & \scalebox{1}{$\left[\begin{array}{c|c|c} S_{y} & S_{x} & - \\ \hline & S_{y} & S_{z} \\ \hline & & S_{y} \\ \end{array}\right]$}\rule[-18pt]{0pt}{40pt} \\ \hline
    $Y_{z}$ & \scalebox{1}{$\begin{bmatrix}\begin{array}{c|c|c} O_{zx}^{x}, O_{yz}^{y}, \bm{O}^{z} & \bm{O}_{lm}^{\alpha} & O_{xy}^{y}, O_{zx}^{z}, \bm{O}^{x} \\ \hline & O_{zx}^{x}, O_{yz}^{y}, \bm{O}^{z} & O_{xy}^{x}, O_{yz}^{z}, \bm{O}^{y} \\ \hline & & O_{zx}^{x}, O_{yz}^{y}, \bm{O}^{z}\end{array}\end{bmatrix}$} & \scalebox{1}{$\left[\begin{array}{c|c|c} S_{z} & - & S_{x} \\ \hline & S_{z} & S_{y} \\ \hline & & S_{z} \\ \end{array}\right]$}\rule[-18pt]{0pt}{40pt} \\ \hline
    $X_{u}$ & \scalebox{1}{$\begin{bmatrix}\begin{array}{c|c|c} \bm{O}_{lm}^{\alpha} & O_{zx}^{x}, O_{yz}^{y}, O_{xx}^{z}, O_{yy}^{z} & O_{xy}^{x}, O_{yz}^{z}, \bm{O}^{y} \\ \hline & \bm{O}_{lm}^{\alpha} & O_{xy}^{y}, O_{zx}^{z}, \bm{O}^{x} \\ \hline & & O_{yz}^{x}, O_{zx}^{y}\end{array}\end{bmatrix}$} & \scalebox{1}{$\left[\begin{array}{c|c|c} - & - & S_{y} \\ \hline & - & S_{x} \\ \hline & & - \\ \end{array}\right]$}\rule[-18pt]{0pt}{40pt} \\ \hline
    $X_{v}$ & \scalebox{1}{$\begin{bmatrix}\begin{array}{c|c|c} \bm{O}_{lm}^{\alpha} & O_{zx}^{x}, O_{yz}^{y}, \bm{O}^{z} & O_{xy}^{x}, O_{yz}^{z}, \bm{O}^{y} \\ \hline & \bm{O}_{lm}^{\alpha} & O_{xy}^{y}, O_{zx}^{z}, \bm{O}^{x} \\ \hline & & \bm{O}_{lm}^{\alpha}\end{array}\end{bmatrix}$} & \scalebox{1}{$\left[\begin{array}{c|c|c} - & S_{z} & S_{y} \\ \hline & - & S_{x} \\ \hline & & - \\ \end{array}\right]$}\rule[-18pt]{0pt}{40pt} \\ \hline
    $X_{yz}$ & \scalebox{1}{$\begin{bmatrix}\begin{array}{c|c|c} O_{yy}^{x}, O_{zz}^{x}, O_{xy}^{y}, O_{zx}^{z} & O_{xy}^{x}, O_{yz}^{z}, \bm{O}^{y} & O_{zx}^{x}, O_{yz}^{y}, \bm{O}^{z} \\ \hline & O_{xy}^{y}, O_{zx}^{z}, \bm{O}^{x} & O_{zx}^{y}, O_{xy}^{z} \\ \hline & & O_{xy}^{y}, O_{zx}^{z}, \bm{O}^{x}\end{array}\end{bmatrix}$} & \scalebox{1}{$\left[\begin{array}{c|c|c} - & S_{y} & S_{z} \\ \hline & S_{x} & - \\ \hline & & S_{x} \\ \end{array}\right]$}\rule[-18pt]{0pt}{40pt} \\ \hline
    $X_{zx}$ & \scalebox{1}{$\begin{bmatrix}\begin{array}{c|c|c} O_{xy}^{x}, O_{yz}^{z}, \bm{O}^{y} & O_{xy}^{y}, O_{zx}^{z}, \bm{O}^{x} & O_{yz}^{x}, O_{xy}^{z} \\ \hline & O_{xy}^{x}, O_{xx}^{y}, O_{zz}^{y}, O_{yz}^{z} & O_{zx}^{x}, O_{yz}^{y}, \bm{O}^{z} \\ \hline & & O_{xy}^{x}, O_{yz}^{z}, \bm{O}^{y}\end{array}\end{bmatrix}$} & \scalebox{1}{$\left[\begin{array}{c|c|c} S_{y} & S_{x} & - \\ \hline & - & S_{z} \\ \hline & & S_{y} \\ \end{array}\right]$}\rule[-18pt]{0pt}{40pt} \\ \hline
    $X_{xy}$ & \scalebox{1}{$\begin{bmatrix}\begin{array}{c|c|c} O_{zx}^{x}, O_{yz}^{y}, \bm{O}^{z} & O_{yz}^{x}, O_{zx}^{y} & O_{xy}^{y}, O_{zx}^{z}, \bm{O}^{x} \\ \hline & O_{zx}^{x}, O_{yz}^{y}, \bm{O}^{z} & O_{xy}^{x}, O_{yz}^{z}, \bm{O}^{y} \\ \hline & & O_{zx}^{x}, O_{yz}^{y}, O_{xx}^{z}, O_{yy}^{z}\end{array}\end{bmatrix}$} & \scalebox{1}{$\left[\begin{array}{c|c|c} S_{z} & - & S_{x} \\ \hline & S_{z} & S_{y} \\ \hline & & - \\ \end{array}\right]$}\rule[-18pt]{0pt}{40pt} \\ \hline
    $Y_{xyz}$ & \scalebox{1}{$\begin{bmatrix}\begin{array}{c|c|c} \bm{O}_{lm}^{\alpha} & O_{zx}^{x}, O_{yz}^{y}, \bm{O}^{z} & O_{xy}^{x}, O_{yz}^{z}, \bm{O}^{y} \\ \hline & \bm{O}_{lm}^{\alpha} & O_{xy}^{y}, O_{zx}^{z}, \bm{O}^{x} \\ \hline & & \bm{O}_{lm}^{\alpha}\end{array}\end{bmatrix}$} & \scalebox{1}{$\left[\begin{array}{c|c|c} - & S_{z} & S_{y} \\ \hline & - & S_{x} \\ \hline & & - \\ \end{array}\right]$}\rule[-18pt]{0pt}{40pt} \\ \hline
    $Y^{\alpha}_{x}$ & \scalebox{1}{$\begin{bmatrix}\begin{array}{c|c|c} O_{xy}^{y}, O_{zx}^{z}, \bm{O}^{x} & O_{xy}^{x}, O_{yz}^{z}, \bm{O}^{y} & O_{zx}^{x}, O_{yz}^{y}, \bm{O}^{z} \\ \hline & O_{xy}^{y}, O_{zx}^{z}, \bm{O}^{x} & \bm{O}_{lm}^{\alpha} \\ \hline & & O_{xy}^{y}, O_{zx}^{z}, \bm{O}^{x}\end{array}\end{bmatrix}$} & \scalebox{1}{$\left[\begin{array}{c|c|c} S_{x} & S_{y} & S_{z} \\ \hline & S_{x} & - \\ \hline & & S_{x} \\ \end{array}\right]$}\rule[-18pt]{0pt}{40pt} \\ \hline
    $Y^{\alpha}_{y}$ & \scalebox{1}{$\begin{bmatrix}\begin{array}{c|c|c} O_{xy}^{x}, O_{yz}^{z}, \bm{O}^{y} & O_{xy}^{y}, O_{zx}^{z}, \bm{O}^{x} & \bm{O}_{lm}^{\alpha} \\ \hline & O_{xy}^{x}, O_{yz}^{z}, \bm{O}^{y} & O_{zx}^{x}, O_{yz}^{y}, \bm{O}^{z} \\ \hline & & O_{xy}^{x}, O_{yz}^{z}, \bm{O}^{y}\end{array}\end{bmatrix}$} & \scalebox{1}{$\left[\begin{array}{c|c|c} S_{y} & S_{x} & - \\ \hline & S_{y} & S_{z} \\ \hline & & S_{y} \\ \end{array}\right]$}\rule[-18pt]{0pt}{40pt} \\ \hline
    $Y^{\alpha}_{z}$ & \scalebox{1}{$\begin{bmatrix}\begin{array}{c|c|c} O_{zx}^{x}, O_{yz}^{y}, \bm{O}^{z} & \bm{O}_{lm}^{\alpha} & O_{xy}^{y}, O_{zx}^{z}, \bm{O}^{x} \\ \hline & O_{zx}^{x}, O_{yz}^{y}, \bm{O}^{z} & O_{xy}^{x}, O_{yz}^{z}, \bm{O}^{y} \\ \hline & & O_{zx}^{x}, O_{yz}^{y}, \bm{O}^{z}\end{array}\end{bmatrix}$} & \scalebox{1}{$\left[\begin{array}{c|c|c} S_{z} & - & S_{x} \\ \hline & S_{z} & S_{y} \\ \hline & & S_{z} \\ \end{array}\right]$}\rule[-18pt]{0pt}{40pt} \\ \hline
    $Y^{\beta}_{x}$ & \scalebox{1}{$\begin{bmatrix}\begin{array}{c|c|c} O_{yy}^{x}, O_{zz}^{x}, O_{xy}^{y}, O_{zx}^{z} & O_{xy}^{x}, O_{yz}^{z}, \bm{O}^{y} & O_{zx}^{x}, O_{yz}^{y}, \bm{O}^{z} \\ \hline & O_{xy}^{y}, O_{zx}^{z}, \bm{O}^{x} & O_{zx}^{y}, O_{xy}^{z} \\ \hline & & O_{xy}^{y}, O_{zx}^{z}, \bm{O}^{x}\end{array}\end{bmatrix}$} & \scalebox{1}{$\left[\begin{array}{c|c|c} - & S_{y} & S_{z} \\ \hline & S_{x} & - \\ \hline & & S_{x} \\ \end{array}\right]$}\rule[-18pt]{0pt}{40pt} \\ \hline
    $Y^{\beta}_{y}$ & \scalebox{1}{$\begin{bmatrix}\begin{array}{c|c|c} O_{xy}^{x}, O_{yz}^{z}, \bm{O}^{y} & O_{xy}^{y}, O_{zx}^{z}, \bm{O}^{x} & O_{yz}^{x}, O_{xy}^{z} \\ \hline & O_{xy}^{x}, O_{xx}^{y}, O_{zz}^{y}, O_{yz}^{z} & O_{zx}^{x}, O_{yz}^{y}, \bm{O}^{z} \\ \hline & & O_{xy}^{x}, O_{yz}^{z}, \bm{O}^{y}\end{array}\end{bmatrix}$} & \scalebox{1}{$\left[\begin{array}{c|c|c} S_{y} & S_{x} & - \\ \hline & - & S_{z} \\ \hline & & S_{y} \\ \end{array}\right]$}\rule[-18pt]{0pt}{40pt} \\ \hline
    $Y^{\beta}_{z}$ & \scalebox{1}{$\begin{bmatrix}\begin{array}{c|c|c} O_{zx}^{x}, O_{yz}^{y}, \bm{O}^{z} & O_{yz}^{x}, O_{zx}^{y} & O_{xy}^{y}, O_{zx}^{z}, \bm{O}^{x} \\ \hline & O_{zx}^{x}, O_{yz}^{y}, \bm{O}^{z} & O_{xy}^{x}, O_{yz}^{z}, \bm{O}^{y} \\ \hline & & O_{zx}^{x}, O_{yz}^{y}, O_{xx}^{z}, O_{yy}^{z}\end{array}\end{bmatrix}$} & \scalebox{1}{$\left[\begin{array}{c|c|c} S_{z} & - & S_{x} \\ \hline & S_{z} & S_{y} \\ \hline & & - \\ \end{array}\right]$}\rule[-18pt]{0pt}{40pt} \\ \hline
  \end{tabular}
  \label{tab:MO spin Ohmic}
\end{table*}

  \begin{table*}
  \caption{\raggedright
    Symmetry-allowed MO Ohmic conductivity tensors with activations of rank-0, -4, and -5 \emph{c}-multipoles (\emph{c}-MP).
    It is noted that a corresponding spin conductivity tensor does not exist.
  }
  \begin{tabular}{c|c||c|c}
    \emph{c}-MP & MO conductivity & \emph{c}-MP & MO conductivity \\ \hline \hline
    $X_{0}$ & \scalebox{0.9}{$\begin{bmatrix}\begin{array}{c|c|c} O_{zx}^{y}, O_{xy}^{z} & O_{zx}^{x}, O_{yz}^{y}, O_{xx}^{z}, O_{yy}^{z} & O_{xy}^{x}, O_{xx}^{y}, O_{zz}^{y}, O_{yz}^{z} \\ \hline & O_{yz}^{x}, O_{xy}^{z} & O_{yy}^{x}, O_{zz}^{x}, O_{xy}^{y}, O_{zx}^{z} \\ \hline & & O_{yz}^{x}, O_{zx}^{y} \\ \end{array}\end{bmatrix}$} & $Y_{5u}$ & \scalebox{0.9}{$\begin{bmatrix}\begin{array}{c|c|c} \bm{O}_{lm}^{\alpha} & O_{zx}^{x}, O_{yz}^{y}, O_{xx}^{z}, O_{yy}^{z} & O_{xy}^{x}, O_{xx}^{y}, O_{yy}^{y} \\ \hline & \bm{O}_{lm}^{\alpha} & O_{xx}^{x}, O_{yy}^{x}, O_{xy}^{y} \\ \hline & & - \\ \end{array}\end{bmatrix}$}\rule[-15pt]{0pt}{35pt} \\ \hline
    $X_{4}$ & \scalebox{0.9}{$\begin{bmatrix}\begin{array}{c|c|c} O_{zx}^{y}, O_{xy}^{z} & O_{zx}^{x}, O_{yz}^{y}, O_{xx}^{z}, O_{yy}^{z} & O_{xy}^{x}, O_{xx}^{y}, O_{zz}^{y}, O_{yz}^{z} \\ \hline & O_{yz}^{x}, O_{xy}^{z} & O_{yy}^{x}, O_{zz}^{x}, O_{xy}^{y}, O_{zx}^{z} \\ \hline & & O_{yz}^{x}, O_{zx}^{y} \\ \end{array}\end{bmatrix}$} & $Y_{5v}$ & \scalebox{0.9}{$\begin{bmatrix}\begin{array}{c|c|c} \bm{O}_{lm}^{\alpha} & O_{zx}^{x}, O_{yz}^{y}, \bm{O}^{z} & O_{xy}^{x}, O_{yz}^{z}, \bm{O}^{y} \\ \hline & \bm{O}_{lm}^{\alpha} & O_{xy}^{y}, O_{zx}^{z}, \bm{O}^{x} \\ \hline & & \bm{O}_{lm}^{\alpha} \\ \end{array}\end{bmatrix}$}\rule[-15pt]{0pt}{35pt} \\ \hline
    $X_{4u}$ & \scalebox{0.9}{$\begin{bmatrix}\begin{array}{c|c|c} \bm{O}_{lm}^{\alpha} & O_{zx}^{x}, O_{yz}^{y}, O_{xx}^{z}, O_{yy}^{z} & O_{xy}^{x}, O_{yz}^{z}, \bm{O}^{y} \\ \hline & \bm{O}_{lm}^{\alpha} & O_{xy}^{y}, O_{zx}^{z}, \bm{O}^{x} \\ \hline & & O_{yz}^{x}, O_{zx}^{y} \\ \end{array}\end{bmatrix}$} & $Y^{\alpha1}_{5x}$ & \scalebox{0.9}{$\begin{bmatrix}\begin{array}{c|c|c} O_{xy}^{y}, O_{zx}^{z}, \bm{O}^{x} & O_{xy}^{x}, O_{yz}^{z}, \bm{O}^{y} & O_{zx}^{x}, O_{yz}^{y}, \bm{O}^{z} \\ \hline & O_{xy}^{y}, O_{zx}^{z}, \bm{O}^{x} & \bm{O}_{lm}^{\alpha} \\ \hline & & O_{xy}^{y}, O_{zx}^{z}, \bm{O}^{x} \\ \end{array}\end{bmatrix}$}\rule[-15pt]{0pt}{35pt} \\ \hline
    $X_{4v}$ & \scalebox{0.9}{$\begin{bmatrix}\begin{array}{c|c|c} \bm{O}_{lm}^{\alpha} & O_{zx}^{x}, O_{yz}^{y}, \bm{O}^{z} & O_{xy}^{x}, O_{yz}^{z}, \bm{O}^{y} \\ \hline & \bm{O}_{lm}^{\alpha} & O_{xy}^{y}, O_{zx}^{z}, \bm{O}^{x} \\ \hline & & \bm{O}_{lm}^{\alpha} \\ \end{array}\end{bmatrix}$} & $Y^{\alpha1}_{5y}$ & \scalebox{0.9}{$\begin{bmatrix}\begin{array}{c|c|c} O_{xy}^{x}, O_{yz}^{z}, \bm{O}^{y} & O_{xy}^{y}, O_{zx}^{z}, \bm{O}^{x} & \bm{O}_{lm}^{\alpha} \\ \hline & O_{xy}^{x}, O_{yz}^{z}, \bm{O}^{y} & O_{zx}^{x}, O_{yz}^{y}, \bm{O}^{z} \\ \hline & & O_{xy}^{x}, O_{yz}^{z}, \bm{O}^{y} \\ \end{array}\end{bmatrix}$}\rule[-15pt]{0pt}{35pt} \\ \hline
    $X^{\alpha}_{4x}$ & \scalebox{0.9}{$\begin{bmatrix}\begin{array}{c|c|c} - & O_{yy}^{y}, O_{zz}^{y}, O_{yz}^{z} & O_{yz}^{y}, O_{yy}^{z}, O_{zz}^{z} \\ \hline & O_{yy}^{x}, O_{zz}^{x}, O_{xy}^{y}, O_{zx}^{z} & \bm{O}_{lm}^{\alpha} \\ \hline & & O_{yy}^{x}, O_{zz}^{x}, O_{xy}^{y}, O_{zx}^{z} \\ \end{array}\end{bmatrix}$} & $Y^{\alpha1}_{5z}$ & \scalebox{0.9}{$\begin{bmatrix}\begin{array}{c|c|c} O_{zx}^{x}, O_{yz}^{y}, \bm{O}^{z} & \bm{O}_{lm}^{\alpha} & O_{xy}^{y}, O_{zx}^{z}, \bm{O}^{x} \\ \hline & O_{zx}^{x}, O_{yz}^{y}, \bm{O}^{z} & O_{xy}^{x}, O_{yz}^{z}, \bm{O}^{y} \\ \hline & & O_{zx}^{x}, O_{yz}^{y}, \bm{O}^{z} \\ \end{array}\end{bmatrix}$}\rule[-15pt]{0pt}{35pt} \\ \hline
    $X^{\alpha}_{4y}$ & \scalebox{0.9}{$\begin{bmatrix}\begin{array}{c|c|c} O_{xy}^{x}, O_{xx}^{y}, O_{zz}^{y}, O_{yz}^{z} & O_{xx}^{x}, O_{zz}^{x}, O_{zx}^{z} & \bm{O}_{lm}^{\alpha} \\ \hline & - & O_{zx}^{x}, O_{xx}^{z}, O_{zz}^{z} \\ \hline & & O_{xy}^{x}, O_{xx}^{y}, O_{zz}^{y}, O_{yz}^{z} \\ \end{array}\end{bmatrix}$} & $Y^{\alpha2}_{5x}$ & \scalebox{0.9}{$\begin{bmatrix}\begin{array}{c|c|c} - & O_{yy}^{y}, O_{zz}^{y}, O_{yz}^{z} & O_{yz}^{y}, O_{yy}^{z}, O_{zz}^{z} \\ \hline & O_{yy}^{x}, O_{zz}^{x}, O_{xy}^{y}, O_{zx}^{z} & \bm{O}_{lm}^{\alpha} \\ \hline & & O_{yy}^{x}, O_{zz}^{x}, O_{xy}^{y}, O_{zx}^{z} \\ \end{array}\end{bmatrix}$}\rule[-15pt]{0pt}{35pt} \\ \hline
    $X^{\alpha}_{4z}$ & \scalebox{0.9}{$\begin{bmatrix}\begin{array}{c|c|c} O_{zx}^{x}, O_{yz}^{y}, O_{xx}^{z}, O_{yy}^{z} & \bm{O}_{lm}^{\alpha} & O_{xx}^{x}, O_{yy}^{x}, O_{xy}^{y} \\ \hline & O_{zx}^{x}, O_{yz}^{y}, O_{xx}^{z}, O_{yy}^{z} & O_{xy}^{x}, O_{xx}^{y}, O_{yy}^{y} \\ \hline & & - \\ \end{array}\end{bmatrix}$} & $Y^{\alpha2}_{5y}$ & \scalebox{0.9}{$\begin{bmatrix}\begin{array}{c|c|c} O_{xy}^{x}, O_{xx}^{y}, O_{zz}^{y}, O_{yz}^{z} & O_{xx}^{x}, O_{zz}^{x}, O_{zx}^{z} & \bm{O}_{lm}^{\alpha} \\ \hline & - & O_{zx}^{x}, O_{xx}^{z}, O_{zz}^{z} \\ \hline & & O_{xy}^{x}, O_{xx}^{y}, O_{zz}^{y}, O_{yz}^{z} \\ \end{array}\end{bmatrix}$}\rule[-15pt]{0pt}{35pt} \\ \hline
    $X^{\beta}_{4x}$ & \scalebox{0.9}{$\begin{bmatrix}\begin{array}{c|c|c} O_{yy}^{x}, O_{zz}^{x}, O_{xy}^{y}, O_{zx}^{z} & O_{xy}^{x}, O_{yz}^{z}, \bm{O}^{y} & O_{zx}^{x}, O_{yz}^{y}, \bm{O}^{z} \\ \hline & O_{xy}^{y}, O_{zx}^{z}, \bm{O}^{x} & O_{zx}^{y}, O_{xy}^{z} \\ \hline & & O_{xy}^{y}, O_{zx}^{z}, \bm{O}^{x} \\ \end{array}\end{bmatrix}$} & $Y^{\alpha2}_{5z}$ & \scalebox{0.9}{$\begin{bmatrix}\begin{array}{c|c|c} O_{zx}^{x}, O_{yz}^{y}, O_{xx}^{z}, O_{yy}^{z} & \bm{O}_{lm}^{\alpha} & O_{xx}^{x}, O_{yy}^{x}, O_{xy}^{y} \\ \hline & O_{zx}^{x}, O_{yz}^{y}, O_{xx}^{z}, O_{yy}^{z} & O_{xy}^{x}, O_{xx}^{y}, O_{yy}^{y} \\ \hline & & - \\ \end{array}\end{bmatrix}$}\rule[-15pt]{0pt}{35pt} \\ \hline
    $X^{\beta}_{4y}$ & \scalebox{0.9}{$\begin{bmatrix}\begin{array}{c|c|c} O_{xy}^{x}, O_{yz}^{z}, \bm{O}^{y} & O_{xy}^{y}, O_{zx}^{z}, \bm{O}^{x} & O_{yz}^{x}, O_{xy}^{z} \\ \hline & O_{xy}^{x}, O_{xx}^{y}, O_{zz}^{y}, O_{yz}^{z} & O_{zx}^{x}, O_{yz}^{y}, \bm{O}^{z} \\ \hline & & O_{xy}^{x}, O_{yz}^{z}, \bm{O}^{y} \\ \end{array}\end{bmatrix}$} & $Y^{\beta}_{5x}$ & \scalebox{0.9}{$\begin{bmatrix}\begin{array}{c|c|c} O_{yy}^{x}, O_{zz}^{x}, O_{xy}^{y}, O_{zx}^{z} & O_{xy}^{x}, O_{xx}^{y}, O_{yy}^{y} & O_{zx}^{x}, O_{xx}^{z}, O_{zz}^{z} \\ \hline & O_{xx}^{x}, O_{yy}^{x}, O_{xy}^{y} & - \\ \hline & & O_{xx}^{x}, O_{zz}^{x}, O_{zx}^{z} \\ \end{array}\end{bmatrix}$}\rule[-15pt]{0pt}{35pt} \\ \hline
    $X^{\beta}_{4z}$ & \scalebox{0.9}{$\begin{bmatrix}\begin{array}{c|c|c} O_{zx}^{x}, O_{yz}^{y}, \bm{O}^{z} & O_{yz}^{x}, O_{zx}^{y} & O_{xy}^{y}, O_{zx}^{z}, \bm{O}^{x} \\ \hline & O_{zx}^{x}, O_{yz}^{y}, \bm{O}^{z} & O_{xy}^{x}, O_{yz}^{z}, \bm{O}^{y} \\ \hline & & O_{zx}^{x}, O_{yz}^{y}, O_{xx}^{z}, O_{yy}^{z} \\ \end{array}\end{bmatrix}$} & $Y^{\beta}_{5y}$ & \scalebox{0.9}{$\begin{bmatrix}\begin{array}{c|c|c} O_{xy}^{x}, O_{xx}^{y}, O_{yy}^{y} & O_{xx}^{x}, O_{yy}^{x}, O_{xy}^{y} & - \\ \hline & O_{xy}^{x}, O_{xx}^{y}, O_{zz}^{y}, O_{yz}^{z} & O_{yz}^{y}, O_{yy}^{z}, O_{zz}^{z} \\ \hline & & O_{yy}^{y}, O_{zz}^{y}, O_{yz}^{z} \\ \end{array}\end{bmatrix}$}\rule[-15pt]{0pt}{35pt} \\ \hline
    & & $Y^{\beta}_{5z}$ & \scalebox{0.9}{$\begin{bmatrix}\begin{array}{c|c|c} O_{zx}^{x}, O_{xx}^{z}, O_{zz}^{z} & - & O_{xx}^{x}, O_{zz}^{x}, O_{zx}^{z} \\ \hline & O_{yz}^{y}, O_{yy}^{z}, O_{zz}^{z} & O_{yy}^{y}, O_{zz}^{y}, O_{yz}^{z} \\ \hline & & O_{zx}^{x}, O_{yz}^{y}, O_{xx}^{z}, O_{yy}^{z} \\ \end{array}\end{bmatrix}$}\rule[-15pt]{0pt}{35pt} \\ \hline
  \end{tabular}
  \label{tab:MO Ohmic}
\end{table*}
  \begin{table*}
  \caption{\raggedright
    Symmetry-allowed MO Hall conductivity tensors and spin Hall conductivity tensors with activations from rank-0 to -2 \emph{c}-multipoles(\emph{c}-MP).
  }
  \centering
  \begin{tabular}{c|c|c}
    \emph{c}-MP & MO conductivity & spin conductivity \\ \hline \hline
    $X_{0}$ & \scalebox{1}{$\begin{bmatrix}\begin{array}{c|c|c} - & O_{zx}^{x}, O_{yz}^{y}, \bm{O}^{z} & O_{xy}^{x}, O_{yz}^{z}, \bm{O}^{y} \\ \hline & - & O_{xy}^{y}, O_{zx}^{z}, \bm{O}^{x} \\ \hline & & - \end{array}\end{bmatrix}$} & \scalebox{1}{$\left[\begin{array}{c|c|c} - & S_{z} & S_{y} \\ \hline & - & S_{x} \\ \hline & & - \\ \end{array}\right]$}\rule[-18pt]{0pt}{40pt}\\ \hline
    $Y_{x}$ & \scalebox{1}{$\begin{bmatrix}\begin{array}{c|c|c} - & O_{xy}^{x}, O_{yz}^{z}, \bm{O}^{y} & O_{zx}^{x}, O_{yz}^{y}, \bm{O}^{z} \\ \hline & - & O_{zx}^{y}, O_{xy}^{z} \\ \hline & & - \end{array}\end{bmatrix}$} & \scalebox{1}{$\left[\begin{array}{c|c|c} - & S_{y} & S_{z} \\ \hline & - & - \\ \hline & & - \\ \end{array}\right]$}\rule[-18pt]{0pt}{40pt}\\ \hline
    $Y_{y}$ & \scalebox{1}{$\begin{bmatrix}\begin{array}{c|c|c} - & O_{xy}^{y}, O_{zx}^{z}, \bm{O}^{x} & O_{yz}^{x}, O_{xy}^{z} \\ \hline & - & O_{zx}^{x}, O_{yz}^{y}, \bm{O}^{z} \\ \hline & & - \end{array}\end{bmatrix}$} & \scalebox{1}{$\left[\begin{array}{c|c|c} - & S_{x} & - \\ \hline & - & S_{z} \\ \hline & & - \\ \end{array}\right]$}\rule[-18pt]{0pt}{40pt}\\ \hline
    $Y_{z}$ & \scalebox{1}{$\begin{bmatrix}\begin{array}{c|c|c} - & O_{yz}^{x}, O_{zx}^{y} & O_{xy}^{y}, O_{zx}^{z}, \bm{O}^{x} \\ \hline & - & O_{xy}^{x}, O_{yz}^{z}, \bm{O}^{y} \\ \hline & & - \end{array}\end{bmatrix}$} & \scalebox{1}{$\left[\begin{array}{c|c|c} - & - & S_{x} \\ \hline & - & S_{y} \\ \hline & & - \\ \end{array}\right]$}\rule[-18pt]{0pt}{40pt}\\ \hline
    $X_{u}$ & \scalebox{1}{$\begin{bmatrix}\begin{array}{c|c|c} - & O_{zx}^{x}, O_{yz}^{y}, \bm{O}^{z} & O_{xy}^{x}, O_{yz}^{z}, \bm{O}^{y} \\ \hline & - & O_{xy}^{y}, O_{zx}^{z}, \bm{O}^{x} \\ \hline & & - \end{array}\end{bmatrix}$} & \scalebox{1}{$\left[\begin{array}{c|c|c} - & S_{z} & S_{y} \\ \hline & - & S_{x} \\ \hline & & - \\ \end{array}\right]$}\rule[-18pt]{0pt}{40pt}\\ \hline
    $X_{v}$ & \scalebox{1}{$\begin{bmatrix}\begin{array}{c|c|c} - & O_{zx}^{x}, O_{yz}^{y}, O_{xx}^{z}, O_{yy}^{z} & O_{xy}^{x}, O_{yz}^{z}, \bm{O}^{y} \\ \hline & - & O_{xy}^{y}, O_{zx}^{z}, \bm{O}^{x} \\ \hline & & - \end{array}\end{bmatrix}$} & \scalebox{1}{$\left[\begin{array}{c|c|c} - & - & S_{y} \\ \hline & - & S_{x} \\ \hline & & - \\ \end{array}\right]$}\rule[-18pt]{0pt}{40pt}\\ \hline
    $X_{yz}$ & \scalebox{1}{$\begin{bmatrix}\begin{array}{c|c|c} - & O_{xy}^{x}, O_{yz}^{z}, \bm{O}^{y} & O_{zx}^{x}, O_{yz}^{y}, \bm{O}^{z} \\ \hline & - & \bm{O}_{lm}^{\alpha} \\ \hline & & - \end{array}\end{bmatrix}$} & \scalebox{1}{$\left[\begin{array}{c|c|c} - & - & S_{x} \\ \hline & - & S_{y} \\ \hline & & - \\ \end{array}\right]$}\rule[-18pt]{0pt}{40pt}\\ \hline
    $X_{zx}$ & \scalebox{1}{$\begin{bmatrix}\begin{array}{c|c|c} - & O_{xy}^{y}, O_{zx}^{z}, \bm{O}^{x} & \bm{O}_{lm}^{\alpha} \\ \hline & - & O_{zx}^{x}, O_{yz}^{y}, \bm{O}^{z} \\ \hline & & - \end{array}\end{bmatrix}$} & \scalebox{1}{$\left[\begin{array}{c|c|c} - & S_{y} & S_{z} \\ \hline & - & - \\ \hline & & - \\ \end{array}\right]$}\rule[-18pt]{0pt}{40pt}\\ \hline
    $X_{xy}$ & \scalebox{1}{$\begin{bmatrix}\begin{array}{c|c|c} - & \bm{O}_{lm}^{\alpha} & O_{xy}^{y}, O_{zx}^{z}, \bm{O}^{x} \\ \hline & - & O_{xy}^{x}, O_{yz}^{z}, \bm{O}^{y} \\ \hline & & - \end{array}\end{bmatrix}$} & \scalebox{1}{$\left[\begin{array}{c|c|c} - & S_{x} & - \\ \hline & - & S_{z} \\ \hline & & - \\ \end{array}\right]$}\rule[-18pt]{0pt}{40pt}\\ \hline
  \end{tabular}
  \label{tab:MO spin Hall}
\end{table*}
  \begin{table*}
  \caption{\raggedright
    Symmetry-allowed MO Hall conductivity tensors with activations of rank-3 and -4 \emph{c}-multipoles (\emph{c}-MP).
    It is noted that a corresponding spin conductivity tensor does not exist.
  }
  \centering
  \begin{tabular}{c|c|c|c}
    \emph{c}-MP & MO conductivity & \emph{c}-MP & MO conductivity \\ \hline \hline
    $Y_{xyz}$ & \scalebox{1}{$\begin{bmatrix}\begin{array}{c|c|c} - & O_{zx}^{x}, O_{yz}^{y}, O_{xx}^{z}, O_{yy}^{z} & O_{xy}^{x}, O_{xx}^{y}, O_{zz}^{y}, O_{yz}^{z} \\ \hline & - & O_{yy}^{x}, O_{zz}^{x}, O_{xy}^{y}, O_{zx}^{z} \\ \hline & & -\end{array}\end{bmatrix}$} & $X_{4}$ & \scalebox{1}{$\begin{bmatrix}\begin{array}{c|c|c} - & O_{zx}^{x}, O_{yz}^{y}, \bm{O}^{z} & O_{xy}^{x}, O_{yz}^{z}, \bm{O}^{y} \\ \hline & - & O_{xy}^{y}, O_{zx}^{z}, \bm{O}^{x} \\ \hline & & -\end{array}\end{bmatrix}$}\rule[-18pt]{0pt}{40pt}\\ \hline
    $Y^{\alpha}_{x}$ & \scalebox{1}{$\begin{bmatrix}\begin{array}{c|c|c} - & O_{xy}^{x}, O_{yz}^{z}, \bm{O}^{y} & O_{zx}^{x}, O_{yz}^{y}, \bm{O}^{z} \\ \hline & - & O_{zx}^{y}, O_{xy}^{z} \\ \hline & & -\end{array}\end{bmatrix}$} & $X_{4u}$ & \scalebox{1}{$\begin{bmatrix}\begin{array}{c|c|c} - & O_{zx}^{x}, O_{yz}^{y}, \bm{O}^{z} & O_{xy}^{x}, O_{yz}^{z}, \bm{O}^{y} \\ \hline & - & O_{xy}^{y}, O_{zx}^{z}, \bm{O}^{x} \\ \hline & & -\end{array}\end{bmatrix}$}\rule[-18pt]{0pt}{40pt}\\ \hline
    $Y^{\alpha}_{y}$ & \scalebox{1}{$\begin{bmatrix}\begin{array}{c|c|c} - & O_{xy}^{y}, O_{zx}^{z}, \bm{O}^{x} & O_{yz}^{x}, O_{xy}^{z} \\ \hline & - & O_{zx}^{x}, O_{yz}^{y}, \bm{O}^{z} \\ \hline & & -\end{array}\end{bmatrix}$} & $X_{4v}$ & \scalebox{1}{$\begin{bmatrix}\begin{array}{c|c|c} - & O_{zx}^{x}, O_{yz}^{y}, O_{xx}^{z}, O_{yy}^{z} & O_{yy}^{y}, O_{zz}^{y}, O_{yz}^{z} \\ \hline & - & O_{xx}^{x}, O_{zz}^{x}, O_{zx}^{z} \\ \hline & & -\end{array}\end{bmatrix}$}\rule[-18pt]{0pt}{40pt}\\ \hline
    $Y^{\alpha}_{z}$ & \scalebox{1}{$\begin{bmatrix}\begin{array}{c|c|c} - & O_{yz}^{x}, O_{zx}^{y} & O_{xy}^{y}, O_{zx}^{z}, \bm{O}^{x} \\ \hline & - & O_{xy}^{x}, O_{yz}^{z}, \bm{O}^{y} \\ \hline & & -\end{array}\end{bmatrix}$} & $X^{\alpha}_{4x}$ & \scalebox{1}{$\begin{bmatrix}\begin{array}{c|c|c} - & O_{yy}^{y}, O_{zz}^{y}, O_{yz}^{z} & O_{yz}^{y}, O_{yy}^{z}, O_{zz}^{z} \\ \hline & - & - \\ \hline & & -\end{array}\end{bmatrix}$}\rule[-18pt]{0pt}{40pt}\\ \hline
    $Y^{\beta}_{x}$ & \scalebox{1}{$\begin{bmatrix}\begin{array}{c|c|c} - & O_{xy}^{x}, O_{yz}^{z}, \bm{O}^{y} & O_{zx}^{x}, O_{yz}^{y}, \bm{O}^{z} \\ \hline & - & \bm{O}_{lm}^{\alpha} \\ \hline & & -\end{array}\end{bmatrix}$} & $X^{\alpha}_{4y}$ & \scalebox{1}{$\begin{bmatrix}\begin{array}{c|c|c} - & O_{xx}^{x}, O_{zz}^{x}, O_{zx}^{z} & - \\ \hline & - & O_{zx}^{x}, O_{xx}^{z}, O_{zz}^{z} \\ \hline & & -\end{array}\end{bmatrix}$}\rule[-18pt]{0pt}{40pt}\\ \hline
    $Y^{\beta}_{y}$ & \scalebox{1}{$\begin{bmatrix}\begin{array}{c|c|c} - & O_{xy}^{y}, O_{zx}^{z}, \bm{O}^{x} & \bm{O}_{lm}^{\alpha} \\ \hline & - & O_{zx}^{x}, O_{yz}^{y}, \bm{O}^{z} \\ \hline & & -\end{array}\end{bmatrix}$} & $X^{\alpha}_{4z}$ & \scalebox{1}{$\begin{bmatrix}\begin{array}{c|c|c} - & - & O_{xx}^{x}, O_{yy}^{x}, O_{xy}^{y} \\ \hline & - & O_{xy}^{x}, O_{xx}^{y}, O_{yy}^{y} \\ \hline & & -\end{array}\end{bmatrix}$}\rule[-18pt]{0pt}{40pt}\\ \hline
    $Y^{\beta}_{z}$ & \scalebox{1}{$\begin{bmatrix}\begin{array}{c|c|c} - & \bm{O}_{lm}^{\alpha} & O_{xy}^{y}, O_{zx}^{z}, \bm{O}^{x} \\ \hline & - & O_{xy}^{x}, O_{yz}^{z}, \bm{O}^{y} \\ \hline & & -\end{array}\end{bmatrix}$} & $X^{\beta}_{4x}$ & \scalebox{1}{$\begin{bmatrix}\begin{array}{c|c|c} - & O_{xy}^{x}, O_{yz}^{z}, \bm{O}^{y} & O_{zx}^{x}, O_{yz}^{y}, \bm{O}^{z} \\ \hline & - & \bm{O}_{lm}^{\alpha} \\ \hline & & -\end{array}\end{bmatrix}$}\rule[-18pt]{0pt}{40pt}\\ \hline
    & & $X^{\beta}_{4y}$ & \scalebox{1}{$\begin{bmatrix}\begin{array}{c|c|c} - & O_{xy}^{y}, O_{zx}^{z}, \bm{O}^{x} & \bm{O}_{lm}^{\alpha} \\ \hline & - & O_{zx}^{x}, O_{yz}^{y}, \bm{O}^{z} \\ \hline & & -\end{array}\end{bmatrix}$}\rule[-18pt]{0pt}{40pt}\\ \hline
    & & $X^{\beta}_{4z}$ & \scalebox{1}{$\begin{bmatrix}\begin{array}{c|c|c} - & \bm{O}_{lm}^{\alpha} & O_{xy}^{y}, O_{zx}^{z}, \bm{O}^{x} \\ \hline & - & O_{xy}^{x}, O_{yz}^{z}, \bm{O}^{y} \\ \hline & & -\end{array}\end{bmatrix}$}\rule[-18pt]{0pt}{40pt}\\ \hline
  \end{tabular}
  \label{tab:MO Hall}
\end{table*}
\end{widetext}

As shown in Tables~\ref{tab:MO spin Ohmic}--\ref{tab:MO Hall}, there are mainly two different points between MO and spin conductivity tensors.
The first difference lies in the rank of the active multipoles. 
In the MO conductivity tensor, multipoles of rank 0 through rank 5 contribute, whereas in the spin conductivity tensor only multipoles of rank 1 through rank 3 are involved. 
This difference manifests itself in the distinct conditions under which each conductivity tensor component can contribute. 
As shown in Table~\ref{tab:MO Ohmic}, multipoles of rank 0, 4, and 5 contribute to the Ohmic-type MO conductivity tensor, but do not contribute to the spin conductivity tensor. Likewise, as summarized in Table~\ref{tab:MO Hall}, multipoles of rank 3 and rank 4 contribute only to the Hall-type MO conductivity tensor. 
Thus, the MO conductivity tensor can appear even in cases where the spin conductivity tensor is forbidden. 

The second difference lies in the nonzero tensor components when the input and output fields are applied in the $ij$ plane. 
For example, in the case of the active dipole $Y_z$, no spin Hall current occurs in the $xy$ plane, while nonzero MO Hall currents with $O^x_{yz}$ and $O^y_{zx}$ are allowed, as shown in Table~\ref{tab:MO spin Hall}. 
In addition, the MO current generally contains spin components that are absent in conventional spin currents. 
For instance, the $X_{0}$ component of $\sigma_{ij}^{\alpha(\mathrm{H})}$ satisfies $J_{i}^{\alpha} = \epsilon_{ij\alpha} \sigma_{ij}^{\alpha (\mathrm{H})} E_{j}$~\cite{murakami2003dissipationless} and carries only the spin-$\alpha$ component, as shown in Table~\ref{tab:MO spin Hall}.
In contrast, the $X_{0}$ component of $\sigma_{ij}^{mn\alpha(\mathrm{H})}$ carries additional spin components that differ from those of the spin current.
These distinctions provide a way of an experimental identification of MO currents and differentiation from spin conductivity.

\subsection{\label{sec:Classification under 32 point groups}Classification under 32 point groups}
We classify the active components of MO Ohmic and Hall conductivities as well as spin conductivities in the representation forms of Eq.~(\ref{eq:nonzero MO conductivity}) and (\ref{eq:nonzero spin conductivity}) under each CPGs except for the triclinic ones ($\bar{1}$ and $1$).
The results are summarized in Table~\ref{tab:active MO and spin conductivity under 32 CPGs}.
This classification is based on the multipole classification under the CPGs given in Table~\ref{tab:active even party multipoles}, where the primary, secondary, and tertiary axes listed in Table~\ref{tab:primary axes} are used.
In these tables, the abbreviation as $\bm{O}_{mn}^{\alpha} = \{O_{yz}^{x}, O_{zx}^{y}, O_{xy}^{z}\}$, $\bm{O}_{yz} = \{O_{yz}^{x}, O_{yz}^{y}, O_{yz}^{z}\}$, $\bm{O}_{zx} = \{O_{zx}^{x}, O_{zx}^{y}, O_{zx}^{z}\}$, $\bm{O}_{xy} = \{O_{xy}^{x}, O_{xy}^{y}, O_{xy}^{z}\}$, $\bm{O}^{i} = \{O_{xx}^{i}, O_{yy}^{i}, O_{zz}^{i}\}$ for $i = x, y, z$, $\mathrm{all} = \{\bm{O}_{yz}, \bm{O}_{zx}, \bm{O}_{xy}, \bm{O}^{x}, \bm{O}^{y}, \bm{O}^{z}\}$ is used for notational simplicity.
It is noted that the leftmost column of Table~\ref{tab:active MO and spin conductivity under 32 CPGs} represents CPGs in which the same components of the MO and spin conductivities become active, rather than those in which the same multipoles become active as in Table~\ref{tab:active even party multipoles}.

As shown in Table~\ref{tab:active MO and spin conductivity under 32 CPGs}, both MO and spin Hall currents in the $ij$-plain are generated under all CPGs.
In contrast, spin Ohmic conductivities are not activated under some CPGs, meaning that the MO Ohmic current can be distinguished from the spin one.
For instance, under $m\bar{3}m$, $432$, and $\bar{4}3m$, only the MO Ohmic conductivities are allowed with the active multipoles $X_{0}$ and $X_{4}$, as shown in Eqs.~(\ref{eq:multipole decomposition of MO Ohmic conductivity}) and (\ref{eq:multipole decomposition of spin Ohmic conductivity}).
\begin{widetext}

\begin{table*}[htbp]
  \caption{\raggedright
    Symmetry-allowed MO and spin conductivity tensors under the CPGs except for the triclinic ones.
    The upper and lower panels represent the results for the Ohmic and Hall conductivities, respectively.
    In this table, the abbreviation as $\bm{S} = \{S_{x}, S_{y}, S_{z}\}$ is used for notational simplicity.
  }
  \centering
  \begin{tabular}{c|c|c}
    CPGs & MO Ohmic conductivity & spin Ohmic conductivity \\ \hline \hline
    $m\bar{3}m$, $432$, $\bar{4}3m$ & \scalebox{0.9}{$\left[\begin{array}{c|c|c} O_{zx}^{y}, O_{xy}^{z} & O_{zx}^{x}, O_{yz}^{y}, O_{xx}^{z}, O_{yy}^{z} & O_{xy}^{x}, O_{xx}^{y}, O_{zz}^{y}, O_{yz}^{z} \\ \hline & O_{yz}^{x}, O_{xy}^{z} & O_{yy}^{x}, O_{zz}^{x}, O_{xy}^{y}, O_{zx}^{z} \\ \hline & & O_{yz}^{x}, O_{zx}^{y} \\ \end{array}\right]$} & --\rule[-15pt]{0pt}{35pt}\\ \hline 
    $23$, $m\bar{3}$ & 
    \multirow{2}{*}{\scalebox{0.9}{$\left[\begin{array}{c|c|c} \bm{O}_{lm}^{\alpha} & O_{zx}^{x}, O_{yz}^{y}, \bm{O}^{z} & O_{xy}^{x}, O_{yz}^{z}, \bm{O}^{y} \\ \hline & \bm{O}_{lm}^{\alpha} & O_{xy}^{y}, O_{zx}^{z}, \bm{O}^{x} \\ \hline & & \bm{O}_{lm}^{\alpha} \\ \end{array}\right]$}} & 
    \multirow{2}{*}{\scalebox{0.9}{$\left[\begin{array}{c|c|c} - & S_{z} & S_{y} \\ \hline & - & S_{x} \\ \hline & & - \\ \end{array}\right]$}}\rule[-8pt]{0pt}{18pt}\\
    $mmm$, $222$, $mm2$ & & \rule[-8pt]{0pt}{18pt}\\ \hline 
    $4/mmm$, $422$ & 
    \multirow{2}{*}{\scalebox{0.9}{$\left[\begin{array}{c|c|c} \bm{O}_{lm}^{\alpha} & O_{zx}^{x}, O_{yz}^{y}, O_{xx}^{z}, O_{yy}^{z} & O_{xy}^{x}, O_{yz}^{z}, \bm{O}^{y} \\ \hline & \bm{O}_{lm}^{\alpha} & O_{xy}^{y}, O_{zx}^{z}, \bm{O}^{x} \\ \hline & & O_{yz}^{x}, O_{zx}^{y} \\ \end{array}\right]$}} & 
    \multirow{2}{*}{\scalebox{0.9}{$\left[\begin{array}{c|c|c} - & - & S_{y} \\ \hline & - & S_{x} \\ \hline & & - \\ \end{array}\right]$}}\rule[-8pt]{0pt}{18pt}\\
    $\bar{4}2m$, $4mm$ & & \rule[-8pt]{0pt}{18pt}\\ \hline 
    $4/m$, $4$, $\bar{4}$ & 
    \scalebox{0.9}{$\left[\begin{array}{c|c|c} O_{zx}^{x}, O_{yz}^{y}, \bm{O}^{z}, \bm{O}_{lm}^{\alpha} & O_{zx}^{x}, O_{yz}^{y}, O_{xx}^{z}, O_{yy}^{z}, \bm{O}_{lm}^{\alpha} & O_{xy}^{x}, O_{xy}^{y}, O_{yz}^{z}, O_{zx}^{z}, \bm{O}^{x}, \bm{O}^{y} \\ \hline & O_{zx}^{x}, O_{yz}^{y}, \bm{O}^{z}, \bm{O}_{lm}^{\alpha} & O_{xy}^{x}, O_{xy}^{y}, O_{yz}^{z}, O_{zx}^{z}, \bm{O}^{x}, \bm{O}^{y} \\ \hline & & O_{yz}^{x}, O_{zx}^{x}, O_{yz}^{y}, O_{zx}^{y}, \bm{O}^{z} \\ \end{array}\right]$} & 
    \scalebox{0.9}{$\left[\begin{array}{c|c|c} S_{z} & - & S_{x}, S_{y} \\ \hline & S_{z} & S_{x}, S_{y} \\ \hline & & S_{z} \\ \end{array}\right]$}\rule[-15pt]{0pt}{35pt}\\ \hline 
    $2/m$, $2$, $m$ & 
    \scalebox{0.9}{$\left[\begin{array}{c|c|c} O_{xy}^{x}, O_{yz}^{z}, \bm{O}^{y}, \bm{O}_{lm}^{\alpha} & O_{zx}^{x}, O_{yz}^{y}, O_{xy}^{y}, O_{zx}^{z}, \bm{O}^{x}, \bm{O}^{z} & O_{xy}^{x}, O_{yz}^{z}, \bm{O}^{y}, \bm{O}_{lm}^{\alpha} \\ \hline & O_{xy}^{x}, O_{yz}^{z}, \bm{O}^{y}, \bm{O}_{lm}^{\alpha} & O_{zx}^{x}, O_{yz}^{y}, O_{xy}^{y}, O_{zx}^{z}, \bm{O}^{x}, \bm{O}^{z} \\ \hline & & O_{xy}^{x}, O_{yz}^{z}, \bm{O}^{y}, \bm{O}_{lm}^{\alpha} \\ \end{array}\right]$} & 
    \scalebox{0.9}{$\left[\begin{array}{c|c|c} S_{y} & S_{x}, S_{z} & S_{y} \\ \hline & S_{y} & S_{x}, S_{z} \\ \hline & & S_{y} \\ \end{array}\right]$}\rule[-15pt]{0pt}{35pt}\\ \hline 
    $6/mmm$, $622$ & 
    \multirow{2}{*}{\scalebox{0.9}{$\left[\begin{array}{c|c|c} \bm{O}_{lm}^{\alpha} & O_{zx}^{x}, O_{yz}^{y}, O_{xx}^{z}, O_{yy}^{z} & O_{xy}^{x}, O_{yz}^{z}, \bm{O}^{y} \\ \hline & \bm{O}_{lm}^{\alpha} & O_{xy}^{y}, O_{zx}^{z}, \bm{O}^{x} \\ \hline & & O_{yz}^{x}, O_{zx}^{y} \\ \end{array}\right]$}} & 
    \multirow{2}{*}{\scalebox{0.9}{$\left[\begin{array}{c|c|c} - & - & S_{y} \\ \hline & - & S_{x} \\ \hline & & - \\ \end{array}\right]$}}\rule[-8pt]{0pt}{18pt}\\
    $\bar{6}m2$, $6mm$ & & \rule[-8pt]{0pt}{18pt}\\ \hline 
    $6/m$, $6$, $\bar{6}$ & 
    \scalebox{0.9}{$\left[\begin{array}{c|c|c} O_{zx}^{x}, O_{yz}^{y}, \bm{O}^{z}, \bm{O}_{lm}^{\alpha} & O_{zx}^{x}, O_{yz}^{y}, O_{xx}^{z}, O_{yy}^{z}, \bm{O}_{lm}^{\alpha} & O_{xy}^{x}, O_{xy}^{y}, O_{yz}^{z}, O_{zx}^{z}, \bm{O}^{x}, \bm{O}^{y} \\ \hline & O_{zx}^{x}, O_{yz}^{y}, \bm{O}^{z}, \bm{O}_{lm}^{\alpha} & O_{xy}^{x}, O_{xy}^{y}, O_{yz}^{z}, O_{zx}^{z}, \bm{O}^{x}, \bm{O}^{y} \\ \hline & & O_{yz}^{x}, O_{zx}^{x}, O_{yz}^{y}, O_{zx}^{y}, \bm{O}^{z} \\ \end{array}\right]$} & 
    \scalebox{0.9}{$\left[\begin{array}{c|c|c} S_{z} & - & S_{x}, S_{y} \\ \hline & S_{z} & S_{x}, S_{y} \\ \hline & & S_{z} \\ \end{array}\right]$}\rule[-15pt]{0pt}{35pt}\\ \hline 
    $\bar{3}m$, $32$, $3m$ & 
    \scalebox{0.9}{$\left[\begin{array}{c|c|c} O_{xy}^{x}, O_{yz}^{z}, \bm{O}^{y}, \bm{O}_{lm}^{\alpha} & O_{zx}^{x}, O_{yz}^{y}, O_{xy}^{y}, O_{xx}^{z}, O_{yy}^{z}, O_{zx}^{z}, \bm{O}^{x} & O_{xy}^{x}, O_{yz}^{z}, \bm{O}^{y}, \bm{O}_{lm}^{\alpha} \\ \hline & O_{xy}^{x}, O_{yz}^{z}, \bm{O}^{y}, \bm{O}_{lm}^{\alpha} & O_{zx}^{x}, O_{yz}^{y}, O_{xy}^{y}, O_{xx}^{z}, O_{yy}^{z}, O_{zx}^{z}, \bm{O}^{x} \\ \hline & & O_{yz}^{x}, O_{xy}^{x}, O_{xx}^{y}, O_{yy}^{y}, O_{zx}^{y} \\ \end{array}\right]$} & 
    \scalebox{0.9}{$\left[\begin{array}{c|c|c} S_{y} & S_{x} & S_{y} \\ \hline & S_{y} & S_{x} \\ \hline & & - \\ \end{array}\right]$}\rule[-15pt]{0pt}{35pt}\\ \hline 
    $\bar{3}$, $3$ & 
    \scalebox{0.9}{$\left[\begin{array}{c|c|c} \mathrm{all} & O_{xx}^{z}, O_{yy}^{z}, \bm{O}^{x}, \bm{O}^{y}, \bm{O}_{xy}, \bm{O}_{yz}, \bm{O}_{zx} & O_{xx}^{z}, O_{yy}^{z}, \bm{O}^{x}, \bm{O}^{y}, \bm{O}_{xy}, \bm{O}_{yz}, \bm{O}_{zx} \\ \hline & \mathrm{all} & O_{xx}^{z}, O_{yy}^{z}, \bm{O}^{x}, \bm{O}^{y}, \bm{O}_{xy}, \bm{O}_{yz}, \bm{O}_{zx} \\ \hline & & O_{xx}^{x}, O_{yy}^{x}, O_{yz}^{x}, O_{zx}^{x}, O_{xy}^{x}, O_{xx}^{y}, O_{yy}^{y}, O_{yz}^{y}, O_{zx}^{y}, O_{xy}^{y}, \bm{O}^{z} \\ \end{array}\right]$} & 
    \scalebox{0.9}{$\left[\begin{array}{c|c|c} \bm{S} & S_{x}, S_{y} & S_{x}, S_{y} \\ \hline & \bm{S} & S_{x}, S_{y} \\ \hline & & S_{z} \\ \end{array}\right]$}\rule[-15pt]{0pt}{35pt}\\ \hline 
    \multicolumn{3}{c}{} \\
    \multicolumn{3}{c}{} \\
    CPG & MO Hall conductivity & spin Hall conductivity \\ \hline \hline
    $m\bar{3}m$, $432$, $\bar{4}3m$ & 
    \multirow{5}{*}{
      \scalebox{0.9}{$\left[\begin{array}{c|c|c} - & O_{zx}^{x}, O_{yz}^{y}, \bm{O}^{z} & O_{xy}^{x}, O_{yz}^{z}, \bm{O}^{y} \\ \hline & - & O_{xy}^{y}, O_{zx}^{z}, \bm{O}^{x} \\ \hline & & - \\ \end{array}\right]$}
    } & 
    \multirow{5}{*}{
      \scalebox{0.9}{$\left[\begin{array}{c|c|c} - & S_{z} & S_{y} \\ \hline & - & S_{x} \\ \hline & & - \\ \end{array}\right]$}
    } \rule[-2pt]{0pt}{12pt} \\
    $23$, $m\bar{3}$ & & \rule[-2pt]{0pt}{12pt} \\
    $4/mmm$, $422$ & & \rule[-2pt]{0pt}{12pt} \\
    $\bar{4}2m$, $4mm$ & & \rule[-2pt]{0pt}{12pt} \\
    $mmm$, $222$, $mm2$ & & \rule[-2pt]{0pt}{12pt} \\ \hline
    $4/m$, $4$, $\bar{4}$ & 
    \scalebox{0.9}{$\left[\begin{array}{c|c|c} - & O_{yz}^{x}, O_{zx}^{x}, O_{yz}^{y}, O_{zx}^{y}, \bm{O}^{z} & O_{xy}^{x}, O_{xy}^{y}, O_{yz}^{z}, O_{zx}^{z}, \bm{O}^{x}, \bm{O}^{y} \\ \hline & - & O_{xy}^{x}, O_{xy}^{y}, O_{yz}^{z}, O_{zx}^{z}, \bm{O}^{x}, \bm{O}^{y} \\ \hline & & - \\ \end{array}\right]$} & 
    \scalebox{0.9}{$\left[\begin{array}{c|c|c} - & S_{z} & S_{x}, S_{y} \\ \hline & - & S_{x}, S_{y} \\ \hline & & - \\ \end{array}\right]$}\rule[-15pt]{0pt}{35pt}\\ \hline 
    $2/m$, $2$, $m$ & 
    \scalebox{0.9}{$\left[\begin{array}{c|c|c} - & O_{zx}^{x}, O_{yz}^{y}, O_{xy}^{y}, O_{zx}^{z}, \bm{O}^{x}, \bm{O}^{z} & O_{xy}^{x}, O_{yz}^{z}, \bm{O}^{y}, \bm{O}_{lm}^{\alpha} \\ \hline & - & O_{zx}^{x}, O_{yz}^{y}, O_{xy}^{y}, O_{zx}^{z}, \bm{O}^{x}, \bm{O}^{z} \\ \hline & & - \\ \end{array}\right]$} & 
    \scalebox{0.9}{$\left[\begin{array}{c|c|c} - & S_{x}, S_{z} & S_{y} \\ \hline & - & S_{x}, S_{z} \\ \hline & & - \\ \end{array}\right]$}\rule[-15pt]{0pt}{35pt}\\ \hline 
    $6/mmm$, $622$ & 
    \multirow{2}{*}{\scalebox{0.9}{$\left[\begin{array}{c|c|c} - & O_{zx}^{x}, O_{yz}^{y}, \bm{O}^{z} & O_{xy}^{x}, O_{yz}^{z}, \bm{O}^{y} \\ \hline & - & O_{xy}^{y}, O_{zx}^{z}, \bm{O}^{x} \\ \hline & & - \\ \end{array}\right]$}} & 
    \multirow{2}{*}{\scalebox{0.9}{$\left[\begin{array}{c|c|c} - & S_{z} & S_{y} \\ \hline & - & S_{x} \\ \hline & & - \\ \end{array}\right]$}}\rule[-8pt]{0pt}{18pt}\\
    $\bar{6}m2$, $6mm$ & & \rule[-8pt]{0pt}{18pt}\\ \hline 
    $6/m$, $6$, $\bar{6}$ & 
    \scalebox{0.9}{$\left[\begin{array}{c|c|c} - & O_{yz}^{x}, O_{zx}^{x}, O_{yz}^{y}, O_{zx}^{y}, \bm{O}^{z} & O_{xy}^{x}, O_{xy}^{y}, O_{yz}^{z}, O_{zx}^{z}, \bm{O}^{x}, \bm{O}^{y} \\ \hline & - & O_{xy}^{x}, O_{xy}^{y}, O_{yz}^{z}, O_{zx}^{z}, \bm{O}^{x}, \bm{O}^{y} \\ \hline & & - \\ \end{array}\right]$} & 
    \scalebox{0.9}{$\left[\begin{array}{c|c|c} - & S_{z} & S_{x}, S_{y} \\ \hline & - & S_{x}, S_{y} \\ \hline & & - \\ \end{array}\right]$}\rule[-15pt]{0pt}{35pt}\\ \hline 
    $\bar{3}m$, $32$, $3m$ & 
    \scalebox{0.9}{$\left[\begin{array}{c|c|c} - & O_{xx}^{x}, O_{yy}^{x}, O_{zx}^{x}, O_{yz}^{y}, O_{xy}^{y}, \bm{O}^{z} & O_{xy}^{x}, O_{yz}^{z}, \bm{O}^{y}, \bm{O}_{lm}^{\alpha} \\ \hline & - & O_{zx}^{x}, O_{yz}^{y}, O_{xy}^{y}, O_{xx}^{z}, O_{yy}^{z}, O_{zx}^{z}, \bm{O}^{x} \\ \hline & & - \\ \end{array}\right]$} & 
    \scalebox{0.9}{$\left[\begin{array}{c|c|c} - & S_{z} & S_{y} \\ \hline & - & S_{x} \\ \hline & & - \\ \end{array}\right]$}\rule[-15pt]{0pt}{35pt}\\ \hline 
    $\bar{3}$, $3$ & 
    \scalebox{0.9}{$\left[\begin{array}{c|c|c} - & O_{xx}^{x}, O_{yy}^{x}, O_{yz}^{x}, O_{zx}^{x}, O_{xy}^{x}, O_{xx}^{y}, O_{yy}^{y}, O_{yz}^{y}, O_{zx}^{y}, O_{xy}^{y}, \bm{O}^{z} & O_{xx}^{z}, O_{yy}^{z}, \bm{O}^{x}, \bm{O}^{y}, \bm{O}_{xy}, \bm{O}_{yz}, \bm{O}_{zx} \\ \hline & - & O_{xx}^{z}, O_{yy}^{z}, \bm{O}^{x}, \bm{O}^{y}, \bm{O}_{xy}, \bm{O}_{yz}, \bm{O}_{zx} \\ \hline & & - \\ \end{array}\right]$} & 
    \scalebox{0.9}{$\left[\begin{array}{c|c|c} - & S_{z} & S_{x}, S_{y} \\ \hline & - & S_{x}, S_{y} \\ \hline & & - \\ \end{array}\right]$}\rule[-15pt]{0pt}{35pt}\\ \hline 
  \end{tabular}
  \label{tab:active MO and spin conductivity under 32 CPGs}
\end{table*}

\begin{table*}
  \caption{\raggedright
    Even-parity multipoles up to rank-5 belonging to a totally symmetric representation in the 32 point groups.
    The independent number of multipole (tensor components) under each point group for MO (spin) Ohmic and Hall conductivities are given in the rightmost part.
    It is noted that all multipoles are activated under the triclinic point groups $\bar{1}, 1$.
  }
  \centering
  \begin{tabular}{|c|cccccc|c|c|} \hline
    \multirow{2}{*}{CPGs} & \multirow{2}{*}{rank 0} & \multirow{2}{*}{rank 1} & \multirow{2}{*}{rank 2} & \multirow{2}{*}{rank 3} & \multirow{2}{*}{rank 4} & \multirow{2}{*}{rank 5} & \multicolumn{2}{c|}{Ohmic / Hall} \\ \cline{8-9}
    & & & & & & & spin & MO \\ \hline \hline
    $m\bar{3}m$, $432$, $\bar{4}3m$ & $X_{0}$ &  &  &  & $X_{4}$ &  & 0 / 1 & 3 / 3\rule[-5pt]{0pt}{14pt}\\
    $23$, $m\bar{3}$ & $X_{0}$ &  &  & $Y_{xyz}$ & $X_{4}$ &  & 1 / 1 & 8 / 5\rule[-5pt]{0pt}{14pt}\\
    $4/mmm$, $422$, $\bar{4}2m$, $4mm$ & $X_{0}$ &  & $X_{u}$ &  & $X_{4}$, $X_{4u}$ &  & 1 / 2 & 10 / 8\rule[-5pt]{0pt}{14pt}\\
    $4/m$, $4$, $\bar{4}$ & $X_{0}$ & $Y_{z}$ & $X_{u}$ & $Y_{z}^{\alpha}$ & $X_{4}$, $X_{4u}$, $X_{4z}^{\alpha}$ & $Y_{5z}^{\alpha1}$, $Y_{5z}^{\alpha2}$ & 4 / 3 & 25 / 14\rule[-5pt]{0pt}{14pt}\\
    $mmm$, $222$, $mm2$ & $X_{0}$ &  & $X_{u}$, $X_{v}$ & $Y_{xyz}$ & $X_{4}$, $X_{4u}$, $X_{4v}$ &  & 3 / 3 & 22 / 15\rule[-5pt]{0pt}{14pt}\\
    $2/m$, $2$, $m$ & $X_{0}$ & $Y_{y}$ & $X_{u}$, $X_{v}$, $X_{zx}$ \rule[-5pt]{0pt}{14pt} & \rule[-5pt]{0pt}{14pt} $Y_{xyz}$, $Y_{y}^{\alpha}$, $Y_{y}^{\beta}$ \rule[-5pt]{0pt}{14pt} & \rule[-5pt]{0pt}{14pt} $X_{4}$, $X_{4u}$, $X_{4v}$, $X_{4y}^{\alpha}$, $X_{4y}^{\beta}$ \rule[-5pt]{0pt}{14pt} & \rule[-5pt]{0pt}{14pt} $Y_{5y}^{\alpha1}$, $Y_{5y}^{\alpha2}$, $Y_{5y}^{\beta}$ \rule[-5pt]{0pt}{14pt} & 8 / 5 & 30 / 28\rule[-5pt]{0pt}{14pt}\\
    $6/mmm$, $622$, $\bar{6}m2$, $6mm$ & $X_{0}$ &  & $X_{u}$ &  & $X_{40}$ &  & 1 / 2 & 8 / 7\rule[-5pt]{0pt}{14pt}\\
    $6/m$, $6$, $\bar{6}$ & $X_{0}$ & $Y_{z}$ & $X_{u}$ & $Y_{z}^{\alpha}$ & $X_{40}$ & $Y_{50}$ & 4 / 3 & 20 / 12\rule[-5pt]{0pt}{14pt}\\
    $\bar{3}m$, $32$, $3m$ & $X_{0}$ &  & $X_{u}$ & $Y_{3b}$ & $X_{40}$, $X_{4b}$ & $Y_{5b}$ & 2 / 2 & 16 / 10\rule[-5pt]{0pt}{14pt}\\
    $\bar{3}$, $3$ & $X_{0}$ & $Y_{z}$ & $X_{u}$ & $Y_{3a}$, $Y_{3b}$, $Y_{z}^{\alpha}$ & $X_{40}$, $X_{4a}$, $X_{4b}$ & $Y_{50}$, $Y_{5a}$, $Y_{5b}$ & 6 / 3 & 36 / 18\rule[-5pt]{0pt}{14pt}\\
    $\bar{1}$, $1$ & all & all & all & all & all & all & 18 / 9 & 108 / 54 \\ \hline
  \end{tabular}
  \label{tab:active even party multipoles}
\end{table*}

  \begin{table}[h]
\caption{\raggedright Primary, secondary, and tertiary axes of seven crystal systems with respect to the symmetry operations in the Cartesian coordinates~\cite{yatsushiro2021multipole}.}
\centering
\begin{tabular}{l|ccc}
 & Primary & Secondary & Tertiary \\ \hline \hline
Cubic        & $\langle100\rangle$ & $\langle111\rangle$ & $\langle110\rangle$ \\
Tetragonal   & [001] & [100] & [110] \\
Orthorhombic & [100] & [010] & [001] \\
Monoclinic   & [010] &  &  \\
Triclinic    &  &  &  \\
Hexagonal    & [001] & [100] & [010] \\
Trigonal     & [001] & [010] &  \\
\hline
\end{tabular}
\label{tab:primary axes}
\end{table}
\end{widetext}

\subsection{\label{sec:Time-reversal property}Time-reversal property}
We have so far assumed the presence of time-reversal symmetry. 
Meanwhile, analogously to spin conductivity~\cite{kimata2019magnetic, Mook_PhysRevResearch.2.023065, chen2021observation, hayami2022spin}, the MO conductivity tensor likewise includes magnetic contributions when time-reversal symmetry is broken. 
In the following, we investigate the role of $\mathcal{T}$ symmetry on the MO conductivity $\sigma_{i; j}^{mn\alpha}$ by analyzing the Kubo formula within the linear response theory~\cite{Watanabe_PhysRevB.96.064432, hayami2018classification,yatsushiro2021multipole}:
\begin{equation}
  \sigma_{i; j}^{mn \alpha}
  =
  -i\frac{e\hbar}{V} \sum_{\xi_{1} \xi_{2}}
  \frac{f(\epsilon_{\xi_{1}}) - f(\epsilon_{\xi_{2}})}{\epsilon_{\xi_{1}} - \epsilon_{\xi_{2}}} 
  \frac{\langle \xi_{1} | J_{j}^{mn\alpha} | \xi_{2} \rangle \langle \xi_{2} | v_{i} | \xi_{1} \rangle}{\epsilon_{\xi_{1}} - \epsilon_{\xi_{2}} + i \delta},
  \label{eq:Kubo formula}
\end{equation}
where $e$, $V$, and $\delta$ denote the elementary charge, the system volume, and the broadening factor, respectively.
$\xi_{i}$ is a set consisting of the band index and wave vector $\bm{k}$, and $\epsilon_{\xi_{i}}$ is the band energy with the eigenvector $\ket{\xi_{i}}$.
$f(\epsilon_{\xi_{i}})$ is the Fermi distribution function,
\begin{equation}
  f(\epsilon_{\xi_{i}}) = \frac{1}{e^{\left(\epsilon_{\xi_{i}} - \mu\right)/k_{\mathrm{B}} T} + 1},
\end{equation}
where $\mu$, $k_{\mathrm{B}}$, and $T$ are the chemical potential, the Boltzmann constant, and the temperature, respectively.
The velocity operator $v_{i}$ is defined as
\begin{equation}
  v_{i} = \frac{1}{\hbar} \frac{\partial H}{\partial k_{i}},
\end{equation}
where $H$ is the tight-bindign Hamiltonian.

The expressoin in Eq.~(\ref{eq:Kubo formula}) is decomposed into two components according to the $\mathcal{T}$ parity:
\begin{equation}
  \begin{split}
    \sigma_{i; j}^{mn \alpha (\mathrm{J})}
    =
    -\frac{e\hbar\delta}{V} \sum_{\xi_{1} \xi_{2}} &
    \frac{f(\epsilon_{\xi_{1}}) - f(\epsilon_{\xi_{2}})}{\epsilon_{\xi_{1}} - \epsilon_{\xi_{2}}} \\
    \times & \frac{\langle \xi_{1} | J_{j}^{mn\alpha} | \xi_{2} \rangle \langle \xi_{2} | v_{i} | \xi_{1} \rangle}{\left(\epsilon_{\xi_{1}} - \epsilon_{\xi_{2}}\right)^{2} + \delta^{2}},
  \end{split}
\end{equation}
\begin{equation}
  \begin{split}
    \sigma_{i; j}^{mn \alpha (\mathrm{E})}
    =
    -i\frac{e\hbar}{V} \sum_{\xi_{1} \xi_{2}}^{\epsilon_{\xi_{1}} \neq \epsilon_{\xi_{2}}} &
    \frac{f(\epsilon_{\xi_{1}}) - f(\epsilon_{\xi_{2}})}{\left(\epsilon_{\xi_{1}} - \epsilon_{\xi_{2}}\right)^{2} + \delta^{2}} \\
    \times & \langle \xi_{1} | J_{j}^{mn\alpha} | \xi_{2} \rangle \langle \xi_{2} | v_{i} | \xi_{1} \rangle,
  \end{split}
\end{equation}
where $\sigma_{i; j}^{mn \alpha (\mathrm{J})}$ generates the dissipative MO current, whose mainly contribution comes from the $\xi_{1} = \xi_{2}$ term and is proportional to $1/\delta$, while $\sigma_{i; j}^{mn \alpha (\mathrm{E})}$ gives rise to dissipationless one, which is independent of $\delta$ in the clean limit of $\delta \to 0$.
Noted that both $\sigma_{i; j}^{mn \alpha (\mathrm{J})}$ and $\sigma_{i; j}^{mn \alpha (\mathrm{E})}$ are real quantities.

By acting the $\mathcal{T}$ operator on $\sigma_{i; j}^{mn \alpha (\mathrm{J/E})}$, one finds that $\sigma_{i; j}^{mn \alpha (\mathrm{J})}$ and $\sigma_{i; j}^{mn \alpha (\mathrm{E})}$ exhibit opposite time-reversal properties:
\begin{align}
  \mathcal{T}(\sigma_{i; j}^{mn \alpha (\mathrm{J})}) & = t_{J_{i}^{mn \alpha}}t_{v_{j}} \sigma_{i; j}^{mn \alpha (\mathrm{J})} \\
  \mathcal{T}(\sigma_{i; j}^{mn \alpha (\mathrm{E})}) & = -t_{J_{i}^{mn \alpha}}t_{v_{j}} \sigma_{i; j}^{mn \alpha (\mathrm{E})},
\end{align}
where $t_{A}$ denotes the time-reversal parity of the operator $A$, defined as $\mathcal{T}(A) = t_{A} A$.
Owing to $t_{J_{i}^{mn \alpha}} = +1$ and $t_{v_{j}} = -1$, $\sigma_{i; j}^{mn \alpha (\mathrm{J})}$ ($\sigma_{i; j}^{mn \alpha (\mathrm{E})}$) is constructed by M and MT (E and ET) multipoles.
Thus, $\sigma_{i; j}^{mn \alpha (\mathrm{J})}$ vanishes in the $\mathcal{T}$-even system, meaning that dissipative MO currents are not allowed under CPGs whose symmetry operations contain $\mathcal{T}$. 

Meanwhile, the dissipative MO currents are symmetry-allowed under gray MPGs and black and white MPGs, where M and/or MT multipoles are activated. 
For example, in the case of the black and white MPG $m\bar{3}m'$, the M octupole $M_{xyz}$ is activated in addition to E multipoles $Q_0$ and $Q_4$~\cite{yatsushiro2021multipole}. 
In this situation, the MO conductivity tensor components associated with $Y_{xyz}$ become nonzero, which arises from the dissipation process. 

\section{\label{sec:Model Analysis}Model Analysis}
As discussed in Sec.~\ref{sec:Symmetry of MO conductivity}, the MO conductivity $\sigma_{i; j}^{mn\alpha}$ becomes nonzero once the corresponding multipoles are activated.
To demonstrate this, we calculate the MO conductivity tensor in the cubic system, where the axial ET octupole $G_{xyz}$ becomes active through the symmetry reduction from $m\bar{3}m$ to $m\bar{3}$.

\subsection{\label{sec:Tight-binding model}Tight-binding model}
We construct a tight-binding Hamiltonian on a three-dimensional simple cubic lattice under the $m\bar{3}m$ symmetry, considering three \emph{p} orbitals $\{p_{x}, p_{y}, p_{z}\}$ and three \emph{f} orbitals $\{f_{x}^{\beta}, f_{y}^{\beta}, f_{z}^{\beta}\}$ at each lattice site:
\begin{equation}
  \mathcal{H} = \sum_{\bm{k}} \sum_{a, a'} \sum_{\sigma, \sigma'} \braket{a\sigma | H(\bm{k}) | a'\sigma'} c_{\bm{k} a \sigma}^{\dagger} c_{\bm{k} a' \sigma'},
  \label{eq:Hamiltonian}
\end{equation}
where $c_{\bm{k} a \sigma}^{\dagger}$ and $c_{\bm{k} a \sigma}$ are the fermionic creation and annihilation operators of the wave number $\bm{k}$, the orbital $a = p_{x}, p_{y}, p_{z}, f_{x}^{\beta}, f_{y}^{\beta}, f_{z}^{\beta}$, and the spin $\sigma = \uparrow, \downarrow$.
The Hamiltonian matrix $H(\bm{k})$ consists of four parts as follows:
\begin{equation}
  H(\bm{k}) = H_{\mathrm{hop}}(\bm{k}) + H_{\mathrm{SOC}} + \Delta_{fp} + g G_{xyz}.
\end{equation}
The first term, $H_{\mathrm{hop}}(\bm{k})$, represents the symmetry-allowed first- and second-nearest-neighbor hopping Hamiltonian based on the Slater--Koster parameters~\cite{slater1954simplified,takegahara1980slater}, which include $\{(pp\sigma), (pp\pi)\}$ for \emph{p} orbitals, $\{(ff\sigma), (ff\pi), (ff\delta), (ff\phi)\}$ for \emph{f} orbitals, and $\{(pf\sigma), (pf\pi)\}$ for the off-diagonal \emph{p}--\emph{f} hybridization.
Here, the second-nearest-neighbor hopping parameters are scaled by a factor of $1/2$, and the explicit forms of the $6 \times 6$ matrices in spinless space are given in Appendix~\ref{app:Hopping Hamiltonian}.
The second term, $H_{\mathrm{SOC}}$, represents the spin-orbit coupling (SOC), given by
\begin{equation}
  H_{\mathrm{SOC}} = \sum_{\mathrm{orb} = p, f} \sum_{i = x, y, z} \lambda_{\mathrm{orb}}\ L_{i} \sigma_{i},
\end{equation}
where 
$\lambda_{\mathrm{orb}}$ for each \emph{p} and \emph{f} orbital (represented by: $\mathrm{orb}$) denotes the strength of the SOC.
$\sigma_{i}$ represents the Pauli matrices and the angular momentum matrices $L_{i}$ are defined by
\begin{equation}
  L_{x} = 
  \begin{pmatrix}
    0 & 0 & 0 \\
    0 & 0 & -i \\
    0 & i & 0
  \end{pmatrix},
  L_{y} = 
  \begin{pmatrix}
    0 & 0 & i  \\
    0 & 0 & 0 \\
    -i & 0 & 0
  \end{pmatrix},
  L_{z} = 
  \begin{pmatrix}
    0 & -i & 0 \\
    i & 0 & 0 \\
    0 & 0 & 0
  \end{pmatrix},
\end{equation}
where the basis functions are taken as $\{p_{x}, p_{y}, p_{z}\}$ or $\{f_{x}^{\beta}, f_{y}^{\beta}, f_{z}^{\beta}\}$.
The third term represents the energy splitting between $p$ and $f$ orbitals, which is given by
\begin{equation}
  \Delta_{pf} = 
  \begin{pmatrix}
    0 & 0 & 0 & 0 & 0 & 0\\
    0 & 0 & 0 & 0 & 0 & 0\\
    0 & 0 & 0 & 0 & 0 & 0\\
    0 & 0 & 0 & \Delta & 0 & 0\\
    0 & 0 & 0 & 0 & \Delta & 0\\
    0 & 0 & 0 & 0 & 0 & \Delta\\
  \end{pmatrix},
\end{equation}
where the basis functions are taken as $\{p_{x}, p_{y}, p_{z}, f_{x}^{\beta}, f_{y}^{\beta}, f_{z}^{\beta}\}$. 
These three terms are symmetry-allowed under the point group $m\bar{3}m$.

To describe the symmetry reduction from $m\bar{3}m$ to $m\bar{3}$, we introduce the symmetry-lowering onsite hybridization as the fourth term in the Hamiltonian, which is given by
\begin{equation}
  g G_{xyz} 
  = 
  \begin{pmatrix}
    0 & 0 & 0 & g & 0 & 0\\
    0 & 0 & 0 & 0 & g & 0\\
    0 & 0 & 0 & 0 & 0 & g\\
    g & 0 & 0 & 0 & 0 & 0\\
    0 & g & 0 & 0 & 0 & 0\\
    0 & 0 & g & 0 & 0 & 0\\
  \end{pmatrix}.
\end{equation}
Here, $G_{xyz}$ represents the ET octupole~\cite{hayami2018microscopic}, and $g$ is the coupling constant. 
When $g \neq 0$, the symmetry of the system reduces to $m\bar{3}$.

\subsection{\label{sec:Numerical results}Numerical Results}
We show the numerical results of the MO conductivity for the Hamiltonian in Eq.~(\ref{eq:Hamiltonian}). 
We especially focus on the Hall contribution $\sigma_{ij}^{mn \alpha (\mathrm{H})}$.
Under the CPG $m\bar{3}$ in Table~\ref{tab:active even party multipoles}, $Q_{0}$, $G_{xyz}$ and $Q_{4}$ are symmetry allowed; no M and MT multipoles are allowed owing to the presence of the $\mathcal{T}$ symmetry.
Thus, we obtain multipole correspondence of $\sigma_{ij}^{mn \alpha (\mathrm{H})}$ by combination Eq.~(\ref{eq:rank-0 Hall}), (\ref{eq:rank-3 Hall}), and (\ref{eq:rank-4 Hall})
\begin{widetext}
  \begin{equation}
    \left(\begin{matrix}
      2 Q^{(1)}_{0} + 8 Q^{(2)}_{0} + 2 Q_{4} & 0 & 0\\
      6 Q^{(2)}_{0} - Q_{4} + 2 G^{(1)}_{xyz} - G^{(2)}_{xyz} & 0 & 0\\
      6 Q^{(2)}_{0} - Q_{4} - 2 G^{(1)}_{xyz} + G^{(2)}_{xyz} & 0 & 0\\
      0 & 0 & 0\\
      0 & 0 & Q^{(1)}_{0} + Q^{(2)}_{0} - Q_{4} - G^{(1)}_{xyz} - G^{(2)}_{xyz}\\
      0 & Q^{(1)}_{0} + Q^{(2)}_{0} - Q_{4} + G^{(1)}_{xyz} + G^{(2)}_{xyz} & 0\\
      0 & 6 Q^{(2)}_{0} - Q_{4} - 2 G^{(1)}_{xyz} + G^{(2)}_{xyz} & 0\\
      0 & 2 Q^{(1)}_{0} + 8 Q^{(2)}_{0} + 2 Q_{4} & 0\\
      0 & 6 Q^{(2)}_{0} - Q_{4} + 2 G^{(1)}_{xyz} - G^{(2)}_{xyz} & 0\\
      0 & 0 & Q^{(1)}_{0} + Q^{(2)}_{0} - Q_{4} + G^{(1)}_{xyz} + G^{(2)}_{xyz}\\
      0 & 0 & 0\\
      Q^{(1)}_{0} + Q^{(2)}_{0} - Q_{4} - G^{(1)}_{xyz} - G^{(2)}_{xyz} & 0 & 0\\
      0 & 0 & 6 Q^{(2)}_{0} - Q_{4} + 2 G^{(1)}_{xyz} - G^{(2)}_{xyz}\\
      0 & 0 & 6 Q^{(2)}_{0} - Q_{4} - 2 G^{(1)}_{xyz} + G^{(2)}_{xyz}\\
      0 & 0 & 2 Q^{(1)}_{0} + 8 Q^{(2)}_{0} + 2 Q_{4}\\
      0 & Q^{(1)}_{0} + Q^{(2)}_{0} - Q_{4} - G^{(1)}_{xyz} - G^{(2)}_{xyz} & 0\\
      Q^{(1)}_{0} + Q^{(2)}_{0} - Q_{4} + G^{(1)}_{xyz} + G^{(2)}_{xyz} & 0 & 0\\
      0 & 0 & 0
    \end{matrix}\right),
    \label{eq:MO conductivity under Th}
  \end{equation}
\end{widetext}
where the different superscripts (1), (2), $\cdots$ represent the same type but independent components in each multipole.
Among the multipoles, the contributions from $X_0$ and $X_4$ are present under high symmetric $m\bar{3}m$, while those from $Y_{xyz}$ are present only under $m\bar{3}$.
Although $(X_0, X_4)$ and $Y_{xyz}$ appears in the same tensor component, we can separate their contributions by taking a linear combination of two tensor components; for example, E multipole components are given by $\sigma_{xy}^{(\mathrm{H})}(Q) \equiv \sigma_{xy}^{yzy (\mathrm{H})} + \sigma_{xy}^{zxx (\mathrm{H})} \leftrightarrow Q_{0} \oplus Q_{0}' \oplus Q_{4}$, while ET multipole components are given by $\sigma_{xy}^{(\mathrm{H})}(G) \equiv \sigma_{xy}^{yzy (\mathrm{H})} - \sigma_{xy}^{zxx (\mathrm{H})} \leftrightarrow G_{xyz} \oplus G_{xyz}'$ from Eq.~(\ref{eq:MO conductivity under Th}).
It is noted that the dissipative contribution $\sigma_{xy}^{(\mathrm{H}/\mathrm{J})}$ vanishes owing to the absence of magnetic-type multipoles.
In the following numerical calculations, we set $e = \hbar = k_{\mathrm{B}} = 1$.
Other parameters are set as shown in Table~\ref{tab:parameters}.

\begin{table}[h]
  \caption{\raggedright Parameter settings in Figure~\ref{fig:vs delta} and \ref{fig:vs g}.}
  \centering
  \begin{tabular}{|c|c||c|c|} \hline
  parameter & value & parameter & value\rule{0pt}{10pt}\\ \hline
  $(pp\sigma)$ & $-1$ & $\lambda_{p}$ & $2$\rule{0pt}{10pt}\\ \hline
  $(pp\pi)$ & $0.4$ & $\lambda_{f}$ & $0.5$\rule{0pt}{10pt}\\ \hline
  $(pf\sigma)$ & $-0.5$ & $\Delta_{pf}$ & $20$\rule{0pt}{10pt}\\ \hline
  $(pf\pi)$ & $0.3$ & $V$ & $2^{15}$\rule{0pt}{10pt}\\ \hline
  $(ff\sigma)$ & $-0.2$ & $T$ & $0.001$\rule{0pt}{10pt}\\ \hline
  $(ff\pi)$ & $0.1$ & $\mu$ & $-2.5$\rule{0pt}{10pt}\\ \hline
  $(ff\delta)$ & $-0.05$ & \multicolumn{2}{c|}{}\rule{0pt}{10pt}\\ \cline{1-2}
  $(ff\phi)$ & $0.01$ & \multicolumn{2}{c|}{}\rule{0pt}{10pt}\\ \hline
  \end{tabular}
  \label{tab:parameters}
\end{table}

\begin{figure}[htbp]
  \centering
  \includegraphics[width=\linewidth]{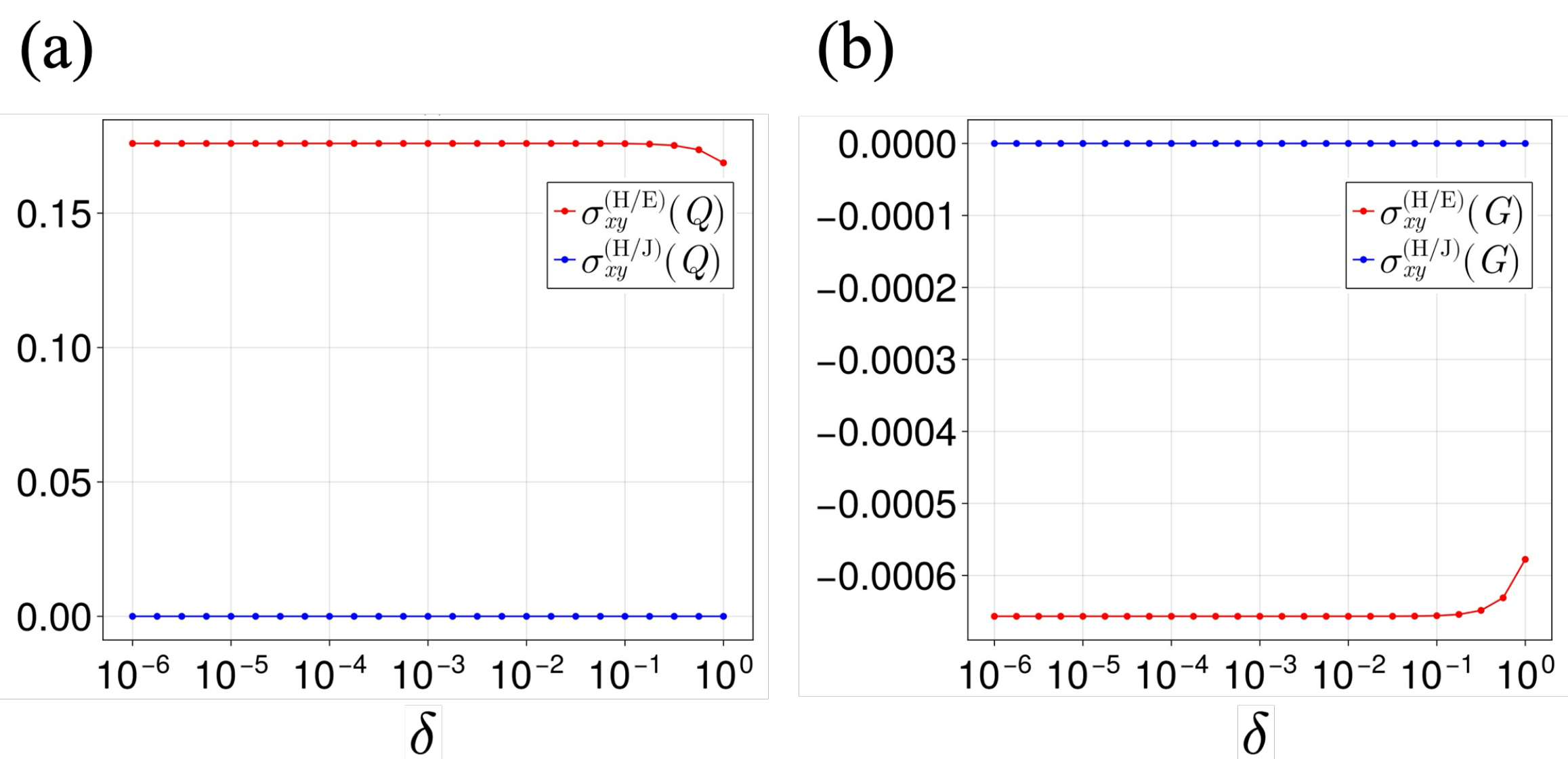}
  \caption{\raggedright 
    The broadening factor $\delta$ dependence of (a) the polar multipole $Q$ components of MO Hall conductivities $\sigma_{xy}^{(\mathrm{H})}(Q)$ and (b) the axial multipole $G$ components of MO Hall conductivities $\sigma_{xy}^{(\mathrm{H})}(G)$.
  }
  \label{fig:vs delta}
\end{figure}

First, we calculate the $\delta$ dependence of $\sigma_{xy}^{(\mathrm{H})}(Q)$ and $\sigma_{xy}^{(\mathrm{H})}(G)$ as shown in Figure~\ref{fig:vs delta}.
The data shows that $\sigma_{xy}^{(\mathrm{H}/\mathrm{J})}(Q, G) = 0$ and $\sigma_{xy}^{(\mathrm{H}/\mathrm{E})}(Q, G) \overset{\delta \to 0}{=} \mathrm{const}$, which is consistent with the discussion in Sec.~\ref{sec:Time-reversal property}.
It results in that $\sigma_{xy}^{(\mathrm{H})}(Q)$ and $\sigma_{xy}^{(\mathrm{H})}(G)$ are constructed from $\mathcal{T}$-even multipoles as discussed in Sec.~\ref{sec:Time-reversal property}, which has no contradiction with the $\mathcal{T}$-even Hamiltonian given in Eq.~(\ref{eq:Hamiltonian}).

\begin{figure}[htbp]
  \centering
  \includegraphics[width=\linewidth]{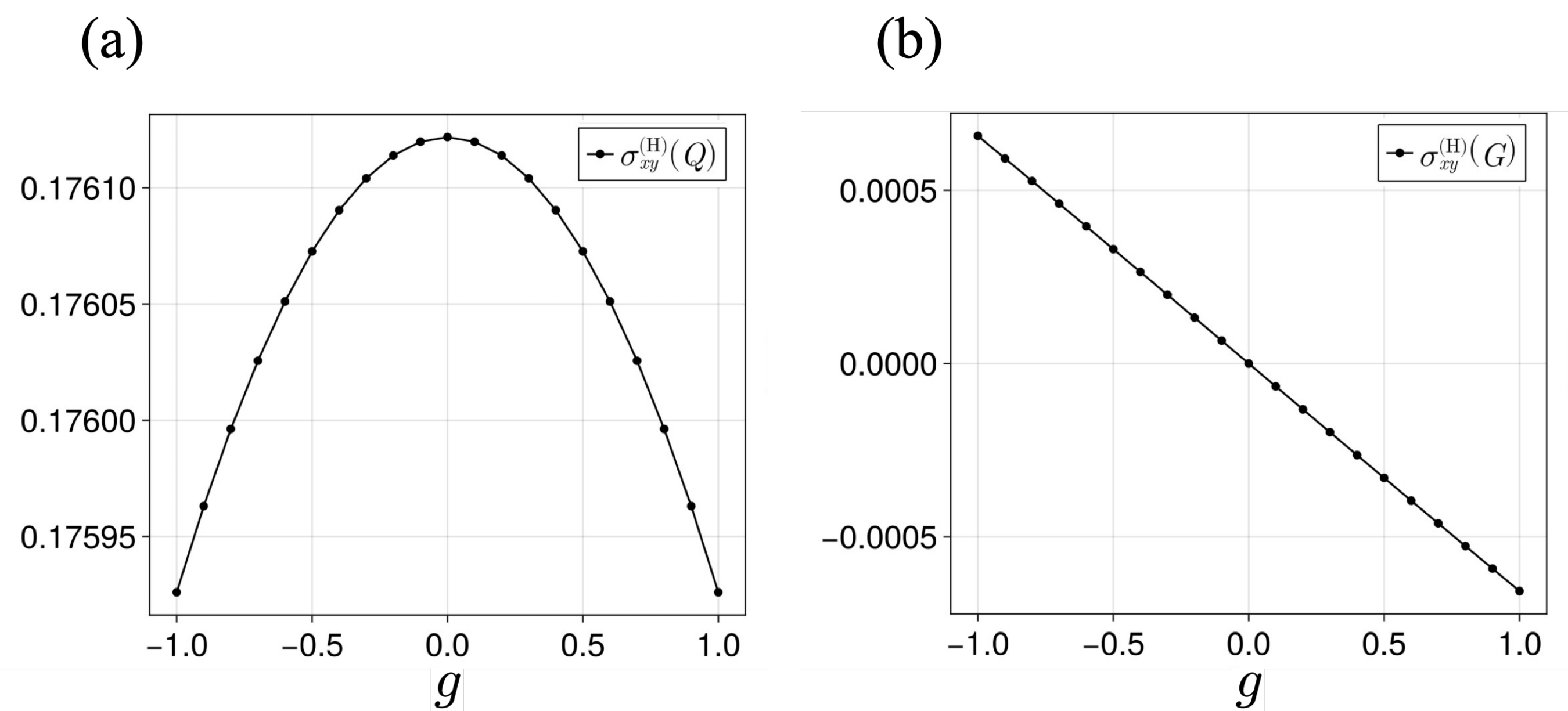}
  \caption{\raggedright 
    The strength of the ETD octupole $g$ dependence of (a) the polar multipole $Q$ components of MO Hall conductivities $\sigma_{xy}^{(\mathrm{H})}(Q)$ and (b) the axial multipole $G$ components of MO Hall conductivities $\sigma_{xy}^{(\mathrm{H})}(G)$.
  }
  \label{fig:vs g}
\end{figure}

Next, we evaluate the contributions from $Q$ and $G$ multipoles by varying the symmetry-lowering parameter $g$.
Figure~\ref{fig:vs g} shows the $g$ dependence of $\sigma_{xy}^{(\mathrm{H})}(Q)$ and $\sigma_{xy}^{(\mathrm{H})}(G)$, demonstrating the activation of $\sigma_{xy}^{(\mathrm{H})}(G)$ due to symmetry lowering induced by the molecular field.
In addition, $\sigma_{xy}^{(\mathrm{H})}(Q)$ expands in even powers of $g$, whereas $\sigma_{xy}^{(\mathrm{H})}(G)$ expands in odd powers, as shown in Fig.~\ref{fig:vs g}.
These results reflect the irreducible representations of $Q(=Q_{0}, Q_{0}', Q_{4})$ and $G(=G_{xyz}, G_{xyz}')$ under the CPG $m\bar{3}m$, which belong to $A_{1g}$ and $A_{2g}$, respectively. 
This feature is also consistent with our symmetry argument in Sec.~\ref{sec:Classification under 32 point groups}.

\section{\label{sec:Summary}Summary}
In summary, we have derived a multipole representation of MO conductivity and provided its classification under 32 crystallographic point groups.
We have shown that the MO conductivity tensor components correspond to polar even-rank and axial odd-rank multipoles, where the multipoles from rank-1 to -5 contribute to the Ohmic contribution, while up to rank 4 contribute to the Hall contribution.
In addition, we have presented the symmetry distinctions between MO and spin currents, which provides a key insight for their experimental identification. 
Furthermore, we have clarified that dissipative MO currents are governed by active E and ET multipoles, whereas nondissipative MO currents are governed by active M and MT multipoles.
Finally, we demonstrate nonzero MO current conductivity by performing the microscopic model calculations. 

Our multipole formulation naturally extends the polar-type rank-5 response tensor---whose spatial parity differs from that of the MO conductivity---to include, for instance, MT-octupole current conductivity via the exchange $X \leftrightarrow Y$.
This extension widens the class of materials capable of exhibiting MO currents, covering not only $d$-wave altermagnets~\cite{ko2025magnetic} but also potential $f$-wave altermagnets and other multipolar systems. 
We expect that this unified framework will guide the exploration and design of octupole current active in a broad range of materials.

\begin{acknowledgments}
  We thank R. Oiwa, Y. Ogawa, and S. Kanda from Hokkaido University for fruitful discussions.
  This research was supported by JSPS KAKENHI Grants Numbers JP21H01037, JP22H00101, JP22H01183, JP23H04869, JP23K03288, JP23K20827, and by JST CREST (JPMJCR23O4) and JST FOREST (JPMJFR2366). 
\end{acknowledgments}

\appendix

\begin{widetext}
  \section{\label{app:Angle dependence of multipole}Angle dependence of multipole}
  We give 
  angle dependence of \emph{c}- and \emph{t}-multipoles in Table~\ref{tab:angle dependences of multipoles}, where angle dependence of \emph{t}-multipole up to rank two is the same as \emph{c}-multipole one.
  
\begin{table*}[htbp]
  \caption{\raggedright Angule dependences of \emph{c}- and \emph{t}-multipoles normalized by $\int_{S} d\Omega X_{i}^{*} X_{j} = \delta_{ij}$ up to rank 5, where $S$, $\Omega$, and $\delta_{ij}$ are the unit sphere, the solid angle, and the Kronecker delta, respectively.
  $r \equiv x^{2} + y^{2} + z^{2}$, and $\mathrm{cyclic}$ denotes the cyclic permutation $x \to y \to z \to x$.}
  \scalebox{0.95}{\begin{minipage}{0.5\linewidth}
    \centering
    \begin{tabular}{ccc}
      rank & \emph{c}-multipole & angle dependence \\ \hline \hline
      $0$ & $X_{0}$ & $\frac{1}{2 \sqrt{\pi}}$ \rule[-8pt]{0pt}{20pt} \\ \hline
      $1$ & $X_{x}, X_{y}, X_{z}$ & $\frac{\sqrt{3} x}{2 \sqrt{\pi} r}, \frac{\sqrt{3} y}{2 \sqrt{\pi} r}, \frac{\sqrt{3} z}{2 \sqrt{\pi} r}$ \rule[-8pt]{0pt}{20pt} \\ \hline
      $2$ & $X_{u}$ & $\frac{\sqrt{5} \left(3 z^{2} - r^{2}\right)}{4 \sqrt{\pi} r^{2}}$ \rule[-8pt]{0pt}{20pt} \\
      & $X_{v}$ & $\frac{\sqrt{15} \left(x^{2} - y^{2}\right)}{4 \sqrt{\pi} r^{2}}$ \rule[-8pt]{0pt}{20pt} \\
      & $X_{yz}, X_{zx}, X_{xy}$ & $\frac{\sqrt{15} y z}{2 \sqrt{\pi} r^{2}}, \text{cyclic}$ \rule[-8pt]{0pt}{20pt} \\ \hline
      $3$ & $X_{xyz}$ & $\frac{\sqrt{105} x y z}{2 \sqrt{\pi} r^{3}}$ \rule[-8pt]{0pt}{20pt} \\
      & $X_{x}^{\alpha}, X_{y}^{\alpha}, X_{z}^{\alpha}$ & $\frac{\sqrt{7} x \left(5 x^{2} - 3 r^{2}\right)}{4 \sqrt{\pi} r^{3}}, \text{cyclic}$ \rule[-8pt]{0pt}{20pt} \\
      & $X_{x}^{\beta}, X_{y}^{\beta}, X_{z}^{\beta}$ & $\frac{\sqrt{105} x \left(y^{2} - z^{2}\right)}{4 \sqrt{\pi} r^{3}}, \text{cyclic}$ \rule[-8pt]{0pt}{20pt} \\ \hline
      $4$ & $X_{4}$ & $\frac{\sqrt{21} \left(x^{4} - 3 x^{2} y^{2} - 3 x^{2} z^{2} + y^{4} - 3 y^{2} z^{2} + z^{4}\right)}{4 \sqrt{\pi} r^{4}}$ \rule[-8pt]{0pt}{20pt} \\
      & $X_{4u}$ & $\frac{\sqrt{15} \left(x^{4} - 12 x^{2} y^{2} + 6 x^{2} z^{2} + y^{4} + 6 y^{2} z^{2} - 2 z^{4}\right)}{8 \sqrt{\pi} r^{4}}$ \rule[-8pt]{0pt}{20pt} \\
      & $X_{4v}$ & $- \frac{3 \sqrt{5} \left(x - y\right) \left(x + y\right) \left(x^{2} + y^{2} - 6 z^{2}\right)}{8 \sqrt{\pi} r^{4}}$ \rule[-8pt]{0pt}{20pt} \\
      & $X_{4x}^{\alpha}, X_{4y}^{\alpha}, X_{4z}^{\alpha}$ & $\frac{3 \sqrt{35} y z \left(y^{2} - z^{2}\right)}{4 \sqrt{\pi} r^{4}}, \text{cyclic}$ \rule[-8pt]{0pt}{20pt} \\
      & $X_{4x}^{\beta}, X_{4y}^{\beta}, X_{4z}^{\beta}$ & $\frac{3 \sqrt{5} y z \left(7 x^{2} - r^{2}\right)}{4 \sqrt{\pi} r^{4}}, \text{cyclic}$ \rule[-8pt]{0pt}{20pt} \\ \hline
      $5$ & $X_{5u}$ & $\frac{3 \sqrt{385} x y z \left(x - y\right) \left(x + y\right)}{4 \sqrt{\pi} r^{5}}$ \rule[-8pt]{0pt}{20pt} \\
      & $X_{5v}$ & $\frac{\sqrt{1155} x y z \left(- x^{2} - y^{2} + 2 z^{2}\right)}{4 \sqrt{\pi} r^{5}}$ \rule[-8pt]{0pt}{20pt} \\
      & $X_{5x}^{\alpha1}, X_{5y}^{\alpha1}, X_{5z}^{\alpha1}$ & $\frac{\sqrt{11} x \left(8 x^{4} - 40 x^{2} \left(y^{2} + z^{2}\right) + 15 \left(y^{2} + z^{2}\right)^{2}\right)}{16 \sqrt{\pi} r^{5}}, \text{cyclic}$ \rule[-8pt]{0pt}{20pt} \\
      & $X_{5x}^{\alpha2}, X_{5y}^{\alpha2}, X_{5z}^{\alpha2}$ & $\frac{3 \sqrt{385} x \left(y^{4} - 6 y^{2} z^{2} + z^{4}\right)}{16 \sqrt{\pi} r^{5}}, \text{cyclic}$ \rule[-8pt]{0pt}{20pt} \\
      & $X_{5x}^{\beta}, X_{5y}^{\beta}, X_{5z}^{\beta}$ & $ \frac{\sqrt{1155} x \left(y^{2} - z^{2}\right) \left(3 z^{2} - r^{2}\right)}{8 \sqrt{\pi} r^{5}}, \text{cyclic}$ \rule[-8pt]{0pt}{20pt} \\ \hline
    \end{tabular}
  \end{minipage}}
  \scalebox{0.95}{\begin{minipage}{0.5\linewidth}
    \begin{tabular}{ccc}
      rank & \emph{t}-multipole & angle dependence \\ \hline \hline
      $3$ & $X_{z}^{\alpha}$ & $\frac{\sqrt{7} z \left(- 3 r^{2} + 5 z^{2}\right)}{4 \sqrt{\pi} r^{3}}$ \rule[-8pt]{0pt}{20pt} \\
      & $X_{3a}$ & $\frac{\sqrt{70} x \left(x^{2} - 3 y^{2}\right)}{8 \sqrt{\pi} r^{3}}$ \rule[-8pt]{0pt}{20pt} \\
      & $X_{3b}$ & $\frac{\sqrt{70} y \left(3 x^{2} - y^{2}\right)}{8 \sqrt{\pi} r^{3}}$ \rule[-8pt]{0pt}{20pt} \\
      & $X_{3u}, X_{3v}$ & $\frac{\sqrt{42} x \left(- r^{2} + 5 z^{2}\right)}{8 \sqrt{\pi} r^{3}}, \frac{\sqrt{42} y \left(- r^{2} + 5 z^{2}\right)}{8 \sqrt{\pi} r^{3}}$ \rule[-8pt]{0pt}{20pt} \\
      & $X_{z}^{\beta}, X_{xyz}$ & $\frac{\sqrt{105} z \left(x^{2} - y^{2}\right)}{4 \sqrt{\pi} r^{3}}, \frac{\sqrt{105} x y z}{2 \sqrt{\pi} r^{3}}$ \rule[-8pt]{0pt}{20pt} \\ \hline
      $4$ & $X_{40}$ & $\frac{3 \left(3 r^{4} - 30 r^{2} z^{2} + 35 z^{4}\right)}{16 \sqrt{\pi} r^{4}}$ \rule[-8pt]{0pt}{20pt} \\
      & $X_{4a}$ & $\frac{3 \sqrt{70} y z \left(3 x^{2} - y^{2}\right)}{8 \sqrt{\pi} r^{4}}$ \rule[-8pt]{0pt}{20pt} \\
      & $X_{4b}$ & $\frac{3 \sqrt{70} x z \left(x^{2} - 3 y^{2}\right)}{8 \sqrt{\pi} r^{4}}$ \rule[-8pt]{0pt}{20pt} \\
      & $X_{4u}^{\alpha}, X_{4v}^{\alpha}$ & $\frac{3 \sqrt{10} x z \left(- 3 r^{2} + 7 z^{2}\right)}{8 \sqrt{\pi} r^{4}}, \frac{3 \sqrt{10} y z \left(- 3 r^{2} + 7 z^{2}\right)}{8 \sqrt{\pi} r^{4}}$ \rule[-8pt]{0pt}{20pt} \\
      & $X_{4u}^{\beta 1}, X_{4v}^{\beta 1}$ & $\frac{3 \sqrt{35} \left(x^{4} - 6 x^{2} y^{2} + y^{4}\right)}{16 \sqrt{\pi} r^{4}}, \frac{3 \sqrt{35} x y \left(x^{2} - y^{2}\right)}{4 \sqrt{\pi} r^{4}}$ \rule[-8pt]{0pt}{20pt} \\
      & $X_{4u}^{\beta 2}, X_{4v}^{\beta 2}$ & $\frac{3 \sqrt{5} \left(x^{2} - y^{2}\right) \left(- r^{2} + 7 z^{2}\right)}{8 \sqrt{\pi} r^{4}}, \frac{3 \sqrt{5} x y \left(- r^{2} + 7 z^{2}\right)}{4 \sqrt{\pi} r^{4}}$ \rule[-8pt]{0pt}{20pt} \\ \hline
      $5$ & $X_{50}$ & $\frac{\sqrt{11} z \left(15 r^{4} - 70 r^{2} z^{2} + 63 z^{4}\right)}{16 \sqrt{\pi} r^{5}}$ \rule[-8pt]{0pt}{20pt} \\
      & $X_{5a}$ & $\frac{\sqrt{770} x \left(x^{2} - 3y^{2}\right) \left(- r^{2} + 9 z^{2}\right)}{32 \sqrt{\pi} r^{5}}$ \rule[-8pt]{0pt}{20pt} \\
      & $X_{5b}$ & $\frac{\sqrt{770} y \left(3x^{2} - y^{2}\right) \left(- r^{2} + 9 z^{2}\right)}{32 \sqrt{\pi} r^{5}}$ \rule[-8pt]{0pt}{20pt} \\
      & $X_{5u}^{\alpha 1}, X_{5v}^{\alpha 1}$ & $\frac{3 \sqrt{154} x \left(x^{4} - 10 x^{2} y^{2} + 5 y^{4}\right)}{32 \sqrt{\pi} r^{5}}, \frac{3 \sqrt{154} y \left(5 x^{4} - 10 x^{2} y^{2} + y^{4}\right)}{32 \sqrt{\pi} r^{5}}$ \rule[-8pt]{0pt}{20pt} \\
      & $X_{5u}^{\alpha 2}, X_{5v}^{\alpha 2}$ & $\frac{\sqrt{165} x \left(r^{4} - 14 r^{2} z^{2} + 21 z^{4}\right)}{16 \sqrt{\pi} r^{5}}, \frac{\sqrt{165} y \left(r^{4} - 14 r^{2} z^{2} + 21 z^{4}\right)}{16 \sqrt{\pi} r^{5}}$ \rule[-8pt]{0pt}{20pt} \\
      & $X_{5u}^{\beta 1}, X_{5v}^{\beta 1}$ & $\frac{3 \sqrt{385} z \left(x^{4} - 6 x^{2} y^{2} + y^{4}\right)}{16 \sqrt{\pi} r^{5}}, \frac{3 \sqrt{385} x y z \left(x^{2} - y^{2}\right)}{4 \sqrt{\pi} r^{5}}$ \rule[-8pt]{0pt}{20pt} \\
      & $X_{5u}^{\beta 2}, X_{5v}^{\beta 2}$ & $\frac{\sqrt{1155} z \left(x^{2} - y^{2}\right) \left(- r^{2} + 3 z^{2}\right)}{8 \sqrt{\pi} r^{5}}, \frac{\sqrt{1155} x y z \left(- r^{2} + 3 z^{2}\right)}{4 \sqrt{\pi} r^{5}}$ \rule[-8pt]{0pt}{20pt} \\ \hline
    \end{tabular}
  \end{minipage}}
  \label{tab:angle dependences of multipoles}
\end{table*}

  \section{\label{app:Multipole expressions of MO conductivities}Multipole expressions of MO conductivities}
  We show multipole expressions of $\sigma_{ij}^{mn \alpha (\mathrm{O, H})}$ up to rank-5 multipoles including \emph{t}-multipoles as well as \emph{c}-multipoles, which expressed as the following form:
  \begin{equation}
    \begin{split}
      & \left(\sigma_{ij}^{mn \alpha (\mathrm{O/L})}\right)^{\top} = \\
      &
        \scalebox{0.65}{$\left(
        \begin{array}{*{18}{c}}
          \sigma_{xx}^{xxx (\mathrm{O})} & \sigma_{xx}^{yyx (\mathrm{O})} & \sigma_{xx}^{zzx (\mathrm{O})} & \sigma_{xx}^{yzx (\mathrm{O})} & \sigma_{xx}^{zxx (\mathrm{O})} & \sigma_{xx}^{xyx (\mathrm{O})} & \sigma_{xx}^{xxy (\mathrm{O})} & \sigma_{xx}^{yyy (\mathrm{O})} & \sigma_{xx}^{zzy (\mathrm{O})} & \sigma_{xx}^{yzy (\mathrm{O})} & \sigma_{xx}^{zxy (\mathrm{O})} & \sigma_{xx}^{xyy (\mathrm{O})} & \sigma_{xx}^{xxz (\mathrm{O})} & \sigma_{xx}^{yyz (\mathrm{O})} & \sigma_{xx}^{zzz (\mathrm{O})} & \sigma_{xx}^{yzz (\mathrm{O})} & \sigma_{xx}^{zxz (\mathrm{O})} & \sigma_{xx}^{xyz (\mathrm{O})} \\
          \sigma_{yy}^{xxx (\mathrm{O})} & \sigma_{yy}^{yyx (\mathrm{O})} & \sigma_{yy}^{zzx (\mathrm{O})} & \sigma_{yy}^{yzx (\mathrm{O})} & \sigma_{yy}^{zxx (\mathrm{O})} & \sigma_{yy}^{xyx (\mathrm{O})} & \sigma_{yy}^{xxy (\mathrm{O})} & \sigma_{yy}^{yyy (\mathrm{O})} & \sigma_{yy}^{zzy (\mathrm{O})} & \sigma_{yy}^{yzy (\mathrm{O})} & \sigma_{yy}^{zxy (\mathrm{O})} & \sigma_{yy}^{xyy (\mathrm{O})} & \sigma_{yy}^{xxz (\mathrm{O})} & \sigma_{yy}^{yyz (\mathrm{O})} & \sigma_{yy}^{zzz (\mathrm{O})} & \sigma_{yy}^{yzz (\mathrm{O})} & \sigma_{yy}^{zxz (\mathrm{O})} & \sigma_{yy}^{xyz (\mathrm{O})} \\
          \sigma_{zz}^{xxx (\mathrm{O})} & \sigma_{zz}^{yyx (\mathrm{O})} & \sigma_{zz}^{zzx (\mathrm{O})} & \sigma_{zz}^{yzx (\mathrm{O})} & \sigma_{zz}^{zxx (\mathrm{O})} & \sigma_{zz}^{xyx (\mathrm{O})} & \sigma_{zz}^{xxy (\mathrm{O})} & \sigma_{zz}^{yyy (\mathrm{O})} & \sigma_{zz}^{zzy (\mathrm{O})} & \sigma_{zz}^{yzy (\mathrm{O})} & \sigma_{zz}^{zxy (\mathrm{O})} & \sigma_{zz}^{xyy (\mathrm{O})} & \sigma_{zz}^{xxz (\mathrm{O})} & \sigma_{zz}^{yyz (\mathrm{O})} & \sigma_{zz}^{zzz (\mathrm{O})} & \sigma_{zz}^{yzz (\mathrm{O})} & \sigma_{zz}^{zxz (\mathrm{O})} & \sigma_{zz}^{xyz (\mathrm{O})} \\
        \end{array}
      \right)$},
    \end{split}
  \end{equation}

  \begin{equation}
    \begin{split}
      & \left(\sigma_{ij}^{mn \alpha (\mathrm{O/T})}\right)^{\top} = \\
      & 
      \scalebox{0.65}{$\left(
        \begin{array}{*{18}{c}}
          \sigma_{yz}^{xxx (\mathrm{O})} & \sigma_{yz}^{yyx (\mathrm{O})} & \sigma_{yz}^{zzx (\mathrm{O})} & \sigma_{yz}^{yzx (\mathrm{O})} & \sigma_{yz}^{zxx (\mathrm{O})} & \sigma_{yz}^{xyx (\mathrm{O})} & \sigma_{yz}^{xxy (\mathrm{O})} & \sigma_{yz}^{yyy (\mathrm{O})} & \sigma_{yz}^{zzy (\mathrm{O})} & \sigma_{yz}^{yzy (\mathrm{O})} & \sigma_{yz}^{zxy (\mathrm{O})} & \sigma_{yz}^{xyy (\mathrm{O})} & \sigma_{yz}^{xxz (\mathrm{O})} & \sigma_{yz}^{yyz (\mathrm{O})} & \sigma_{yz}^{zzz (\mathrm{O})} & \sigma_{yz}^{yzz (\mathrm{O})} & \sigma_{yz}^{zxz (\mathrm{O})} & \sigma_{yz}^{xyz (\mathrm{O})} \\
          \sigma_{zx}^{xxx (\mathrm{O})} & \sigma_{zx}^{yyx (\mathrm{O})} & \sigma_{zx}^{zzx (\mathrm{O})} & \sigma_{zx}^{yzx (\mathrm{O})} & \sigma_{zx}^{zxx (\mathrm{O})} & \sigma_{zx}^{xyx (\mathrm{O})} & \sigma_{zx}^{xxy (\mathrm{O})} & \sigma_{zx}^{yyy (\mathrm{O})} & \sigma_{zx}^{zzy (\mathrm{O})} & \sigma_{zx}^{yzy (\mathrm{O})} & \sigma_{zx}^{zxy (\mathrm{O})} & \sigma_{zx}^{xyy (\mathrm{O})} & \sigma_{zx}^{xxz (\mathrm{O})} & \sigma_{zx}^{yyz (\mathrm{O})} & \sigma_{zx}^{zzz (\mathrm{O})} & \sigma_{zx}^{yzz (\mathrm{O})} & \sigma_{zx}^{zxz (\mathrm{O})} & \sigma_{zx}^{xyz (\mathrm{O})} \\
          \sigma_{xy}^{xxx (\mathrm{O})} & \sigma_{xy}^{yyx (\mathrm{O})} & \sigma_{xy}^{zzx (\mathrm{O})} & \sigma_{xy}^{yzx (\mathrm{O})} & \sigma_{xy}^{zxx (\mathrm{O})} & \sigma_{xy}^{xyx (\mathrm{O})} & \sigma_{xy}^{xxy (\mathrm{O})} & \sigma_{xy}^{yyy (\mathrm{O})} & \sigma_{xy}^{zzy (\mathrm{O})} & \sigma_{xy}^{yzy (\mathrm{O})} & \sigma_{xy}^{zxy (\mathrm{O})} & \sigma_{xy}^{xyy (\mathrm{O})} & \sigma_{xy}^{xxz (\mathrm{O})} & \sigma_{xy}^{yyz (\mathrm{O})} & \sigma_{xy}^{zzz (\mathrm{O})} & \sigma_{xy}^{yzz (\mathrm{O})} & \sigma_{xy}^{zxz (\mathrm{O})} & \sigma_{xy}^{xyz (\mathrm{O})} \\
        \end{array}
      \right)$}.
    \end{split}
  \end{equation}
  Here, $\sigma_{ij}^{mn \alpha (\mathrm{O})}$ is decomposed into the longitudinal part $\sigma_{ij}^{mn \alpha (\mathrm{O/L})}$ with $i=j$ and the transverse part $\sigma_{ij}^{mn \alpha (\mathrm{O/T})}$ with $i \neq j$, and $\sigma_{ij}^{mn \alpha (\mathrm{H})}$ is shown in Eq.~(\ref{eq:index table of MOHE}).
  We aditionally show the normalized MO conductivity $\tilde{\sigma}_{ij}^{mn \alpha (\mathrm{O, H})}$ for better readability, whose multipole do not obey Eq.~(\ref{eq:multipole decomposition of MO conductivities with CG}) but have the same irreducible representations as one of $\tilde{\sigma}_{ij}^{mn \alpha (\mathrm{O, H})}$ in arbitary point groups.
  The independent multipole degrees of freedom are distinguished by the superscript $(l)$ for the integer $l$.
  
  \subsection{rank-0 \emph{c}-multipoles}
  \begin{align}
  \nonumber
  & \left(\sigma_{ij}^{mn\alpha(\mathrm{O/L})}\right)^{\top} =  \\
  & \left(
\begin{array}{cccccccccccccccccc}0 & 0 & 0 & 0 & 0 & 0 & 0 & 0 & 0 & 0 & - X_{0} & 0 & 0 & 0 & 0 & 0 & 0 & X_{0}\\0 & 0 & 0 & X_{0} & 0 & 0 & 0 & 0 & 0 & 0 & 0 & 0 & 0 & 0 & 0 & 0 & 0 & - X_{0}\\0 & 0 & 0 & - X_{0} & 0 & 0 & 0 & 0 & 0 & 0 & X_{0} & 0 & 0 & 0 & 0 & 0 & 0 & 0\end{array}\right).
\end{align}
\begin{align}
  \nonumber
  & \left(\sigma_{ij}^{mn\alpha(\mathrm{O/T})}\right)^{\top} =  \\
  & \left(\begin{array}{cccccccccccccccccc}0 & - X_{0} & X_{0} & 0 & 0 & 0 & 0 & 0 & 0 & 0 & 0 & \frac{X_{0}}{2} & 0 & 0 & 0 & 0 & - \frac{X_{0}}{2} & 0\\0 & 0 & 0 & 0 & 0 & - \frac{X_{0}}{2} & X_{0} & 0 & - X_{0} & 0 & 0 & 0 & 0 & 0 & 0 & \frac{X_{0}}{2} & 0 & 0\\0 & 0 & 0 & 0 & \frac{X_{0}}{2} & 0 & 0 & 0 & 0 & - \frac{X_{0}}{2} & 0 & 0 & - X_{0} & X_{0} & 0 & 0 & 0 & 0\end{array}%
\right).
\end{align}

  \subsection{rank-1 \emph{c}-multipoles}
  \begin{align}
  & \sigma_{ij}^{mn\alpha(\mathrm{O/L})} = \sigma_{1}^{(\mathrm{O/L})} + \sigma_{2}^{(\mathrm{O/L})},\\
  \nonumber
  & \sigma_{1}^{(\mathrm{O/L})} = \\
  & \left(%
\right),
\end{align}
where the multipole components with superscript (3) and (6) of $\sigma_{ij}^{mn\alpha(\mathrm{O/T})}$ do not appear.
  \begin{align}
  \nonumber
  & \sigma_{ij}^{mn\alpha(\mathrm{H})} = \\
  & \left(%
\right).
\end{align}

  \subsection{rank-2 \emph{c}-multipoles}
  \begin{align}
  & \sigma_{ij}^{mn\alpha(\mathrm{O/L})} = \sigma_{1}^{(\mathrm{O/L})} + \sigma_{2}^{(\mathrm{O/L})}, \\
  \nonumber
  & \sigma_{1}^{(\mathrm{O/L})} = \\
  & \left(%
\right)$},
\end{align}
where the multipole component with superscript (3) of $\sigma_{ij}^{mn\alpha(\mathrm{O/T})}$ does not appear.
  \begin{align}
  & \sigma_{ij}^{mn\alpha(\mathrm{H})} = \sigma_{1}^{(\mathrm{H})} + \sigma_{2}^{(\mathrm{H})}, \\
  \nonumber
  & \sigma_{1}^{(\mathrm{H})} = \\
  & \left(%
\right)
  $}.
\end{align}

  \subsection{rank-3 \emph{c}-multipoles}
  \begin{align}
  & \sigma_{ij}^{mn\alpha(\mathrm{O/L})} = \sigma_{1}^{(\mathrm{O/L})} + \sigma_{2}^{(\mathrm{O/L})} + \sigma_{3}^{(\mathrm{O/L})}, \\
  \nonumber
  & \sigma_{1}^{(\mathrm{O/L})} = \\
  & \scalebox{0.7}{$\left(%
\right)$},
\end{align}
where the multipole component with superscript (5) of $\sigma_{ij}^{mn\alpha(\mathrm{O/T})}$ does not appear.

  \begin{align}
  \nonumber
  & \sigma_{ij}^{mn\alpha(\mathrm{H})} = \\
  & \scalebox{0.7}{$
    \left(%
\right)
  $}.
\end{align}

  \subsection{rank-4 \emph{c}-multipoles}
  \begin{align}
  & \sigma_{ij}^{mn\alpha(\mathrm{O/L})} = \sigma_{1}^{(\mathrm{O/L})} + \sigma_{2}^{(\mathrm{O/L})}, \\
  \nonumber
  & \sigma_{1}^{(\mathrm{O/L})} = \\
  & \scalebox{1}{$\left(%
\right).
\end{align}

  \subsection{rank-5 \emph{c}-multipoles}
  \begin{align}
  \nonumber
  & \sigma_{ij}^{mn\alpha(\mathrm{O/L})} \\
  & \left(%
\right).
\end{align}

  \subsection{rank-3 \emph{t}-multipoles}
  \begin{align}
  & \sigma_{ij}^{mn\alpha(\mathrm{O/L})} = \sigma_{1}^{(\mathrm{O/L})} + \sigma_{2}^{(\mathrm{O/L})} + \sigma_{3}^{(\mathrm{O/L})}, \\
  \nonumber
  & \sigma_{1}^{(\mathrm{O/L})} = \\
  & \scalebox{0.7}{$\left(%
\right)$},
\end{align}
where the multipole component with superscript (5) of $\sigma_{ij}^{mn\alpha(\mathrm{O/T})}$ does not appear.
  \begin{align}
  \nonumber
  & \sigma_{ij}^{mn\alpha(\mathrm{H})} = \\
  & \scalebox{0.8}{$
    \left(%
\right)
  $}.
\end{align}

  \subsection{rank-4 \emph{t}-multipoles}
  \begin{align}
  & \sigma_{ij}^{mn\alpha(\mathrm{O/L})} = \sigma_{1}^{(\mathrm{O/L})} + \sigma_{2}^{(\mathrm{O/L})}, \\
  \nonumber
  & \sigma_{1}^{(\mathrm{O/L})} = \\
  & \scalebox{1}{$\left(%
\right).
\end{align}

  \subsection{rank-5 \emph{t}-multipoles}
  \begin{align}
  \nonumber
  & \sigma_{ij}^{mn\alpha(\mathrm{O/L})} = \\
  & \left[%
\right].
\end{align}

  \subsection{rank-0 \emph{c}-multipoles with normalization}
  \begin{align}
  \nonumber
  & \left(\tilde{\sigma}_{ij}^{mn\alpha(\mathrm{O/L})}\right)^{\top} = \\
  & \left(\begin{array}{cccccccccccccccccc}0 & 0 & 0 & 0 & 0 & 0 & 0 & 0 & 0 & 0 & - 2 X_{0} & 0 & 0 & 0 & 0 & 0 & 0 & 2 X_{0}\\0 & 0 & 0 & 2 X_{0} & 0 & 0 & 0 & 0 & 0 & 0 & 0 & 0 & 0 & 0 & 0 & 0 & 0 & - 2 X_{0}\\0 & 0 & 0 & - 2 X_{0} & 0 & 0 & 0 & 0 & 0 & 0 & 2 X_{0} & 0 & 0 & 0 & 0 & 0 & 0 & 0\end{array}\right).
\end{align}
\begin{align}
  \nonumber
  & \left(\tilde{\sigma}_{ij}^{mn\alpha(\mathrm{O/T})}\right)^{\top} = \\
  & \left(\begin{array}{cccccccccccccccccc}0 & - 2 X_{0} & 2 X_{0} & 0 & 0 & 0 & 0 & 0 & 0 & 0 & 0 & X_{0} & 0 & 0 & 0 & 0 & - X_{0} & 0\\0 & 0 & 0 & 0 & 0 & - X_{0} & 2 X_{0} & 0 & - 2 X_{0} & 0 & 0 & 0 & 0 & 0 & 0 & X_{0} & 0 & 0\\0 & 0 & 0 & 0 & X_{0} & 0 & 0 & 0 & 0 & - X_{0} & 0 & 0 & - 2 X_{0} & 2 X_{0} & 0 & 0 & 0 & 0\end{array}%
\right).
  \label{eq:rank-0 Hall}
\end{align}

  \subsection{rank-1 \emph{c}-multipoles with normalization}
  \begin{align}
  & \tilde{\sigma}_{ij}^{mn\alpha(\mathrm{O/L})} = \tilde{\sigma}_{1}^{(\mathrm{O/L})} + \tilde{\sigma}_{2}^{(\mathrm{O/L})}, \\
  \nonumber
  & \tilde{\sigma}_{1}^{(\mathrm{O/L})} = \\
  & \left(%
\right),
\end{align}
where the multipole components with superscript (3) and (6) of $\tilde{\sigma}_{ij}^{mn\alpha(\mathrm{O/T})}$ do not appear.
  \begin{align}
  \nonumber
  & \tilde{\sigma}_{ij}^{mn\alpha(\mathrm{H})} = \\
  & \left(%
\right).
\end{align}

  \subsection{rank-2 \emph{c}-multipoles with normalization}
  \begin{align}
  & \tilde{\sigma}_{ij}^{mn\alpha(\mathrm{O/L})} = \tilde{\sigma}_{1}^{(\mathrm{O/L})} + \tilde{\sigma}_{2}^{(\mathrm{O/L})}, \\
  \nonumber
  & \tilde{\sigma}_{1}^{(\mathrm{O/L})} = \\
  & \left(%
\right)$},
\end{align}
where the multipole component with superscript (3) of $\tilde{\sigma}_{ij}^{mn\alpha(\mathrm{O/T})}$ does not appear.
  \begin{align}
  \nonumber
  & \tilde{\sigma}_{ij}^{mn\alpha(\mathrm{H})} = \tilde{\sigma}_{1}^{(\mathrm{H})} + \tilde{\sigma}_{2}^{(\mathrm{H})}, \\
  \nonumber
  & \tilde{\sigma}_{1}^{(\mathrm{H})} = \\
  & \left(%
\right)
\end{align}

  \subsection{rank-3 \emph{c}-multipoles with normalization}
  \begin{align}
  & \tilde{\sigma}_{ij}^{mn\alpha(\mathrm{O/L})} = \tilde{\sigma}_{1}^{(\mathrm{O/L})} + \tilde{\sigma}_{2}^{(\mathrm{O/L})} + \tilde{\sigma}_{3}^{(\mathrm{O/L})}, \\
  \nonumber
  & \tilde{\sigma}_{1}^{(\mathrm{O/L})} = \\
  & \scalebox{0.8}{$\left(%
\right)$},
\end{align}
where the multipole component with superscript (5) of $\tilde{\sigma}_{ij}^{mn\alpha(\mathrm{O/T})}$ does not appear.
  \begin{align}
  \nonumber
  & \tilde{\sigma}_{ij}^{mn\alpha(\mathrm{H})} = \\
  & \left(%
\right).
  \label{eq:rank-3 Hall}
\end{align}

  \subsection{rank-4 \emph{c}-multipoles with normalization}
  \begin{align}
  & \tilde{\sigma}_{ij}^{mn\alpha(\mathrm{O/L})} = \tilde{\sigma}_{1}^{(\mathrm{O/L})} + \tilde{\sigma}_{2}^{(\mathrm{O/L})}, \\
  \nonumber
  & \tilde{\sigma}_{1}^{(\mathrm{O/L})} = \\
  & \left(%
\right).
  \label{eq:rank-4 Hall}
\end{align}

  \subsection{rank-5 \emph{c}-multipoles with normalization}
  \begin{align}
  \nonumber
  & \tilde{\sigma}_{ij}^{mn\alpha(\mathrm{O/L})} \\
  & \left(%
\right).
\end{align}

  \subsection{rank-3 \emph{t}-multipoles with normalization}
  \begin{align}
  & \tilde{\sigma}_{ij}^{mn\alpha(\mathrm{O/L})} = \tilde{\sigma}_{1}^{(\mathrm{O/L})} + \tilde{\sigma}_{2}^{(\mathrm{O/L})} + \tilde{\sigma}_{3}^{(\mathrm{O/L})}, \\
  \nonumber
  & \tilde{\sigma}_{1}^{(\mathrm{O/L})} = \\
  & \scalebox{0.7}{$\left(%
\right)$},
\end{align}
where the multipole component with superscript (5) of $\tilde{\sigma}_{ij}^{mn\alpha(\mathrm{O/T})}$ does not appear.
  \begin{align}
  \nonumber
  & \tilde{\sigma}_{ij}^{mn\alpha(\mathrm{H})} = \\
  & \left(%
\right).
\end{align}

  \subsection{rank-4 \emph{t}-multipoles with normalization}
  \begin{align}
  & \tilde{\sigma}_{ij}^{mn\alpha(\mathrm{O/L})} = \tilde{\sigma}_{1}^{(\mathrm{O/L})} + \tilde{\sigma}_{2}^{(\mathrm{O/L})}, \\
  \nonumber
  & \tilde{\sigma}_{1}^{(\mathrm{O/L})} = \\
  & \scalebox{1}{$\left(%
\right).
\end{align}

  \subsection{rank-5 \emph{t}-multipoles with normalization}
  \begin{align}
  \nonumber
  & \tilde{\sigma}_{ij}^{mn\alpha(\mathrm{O/L})} = \\
  & \left(%
\right).
\end{align}

  \section{\label{app:Comparison between the magnetic octupole conductivities and spin conductivities for t-multipoles}Comparison between the magnetic octupole conductivities and spin conductivities \\for \emph{t}-multipoles}
  We give nonvanishing tensor components of MO and spin conductivities with activations of \emph{t}-multipoles corresponding to Tables~\ref{tab:MO spin Ohmic tesseral}, \ref{tab:MO Ohmic tesseral}, and \ref{tab:MO Hall tesseral}.
  As shown in Tables~\ref{tab:MO spin Ohmic tesseral}--\ref{tab:MO Hall tesseral}, the conclusions discussed in Sec.~\ref{sec:Comparison between the magnetic octupole and spin conductivities}---namely,
  \begin{itemize}
    \item certain multipoles generate only the MO currents,
    \item the MO conductivities carry different spin components from the spin conductivities, and
    \item the $ij$-plain in which the MO or spin currents flow differs even for the same multipole component,
  \end{itemize}
  do not depend on whether the \emph{c}- or \emph{t}-multipoles are used.
  \begin{table*}
  \caption{\raggedright
    Symmetry-allowed MO Ohmic conductivity tensors and spin Ohmic conductivity tensors with activations rank-3 \emph{t}-multipoles(\emph{t}-MP).
  }
  \centering
\begin{tabular}{c|c|c}
    \emph{t}-MP & MO conductivity & spin conductivity \\ \hline \hline
    $Y^{\alpha}_{z}$ & \scalebox{1}{$\begin{bmatrix}\begin{array}{c|c|c} O_{zx}^{x}, O_{yz}^{y}, \bm{O}^{z} & \bm{O}_{lm}^{\alpha} & O_{xy}^{y}, O_{zx}^{z}, \bm{O}^{x} \\ \hline & O_{zx}^{x}, O_{yz}^{y}, \bm{O}^{z} & O_{xy}^{x}, O_{yz}^{z}, \bm{O}^{y} \\ \hline & & O_{zx}^{x}, O_{yz}^{y}, \bm{O}^{z}\end{array}\end{bmatrix}$} & \scalebox{1}{$\left[\begin{array}{c|c|c} S_{z} & - & S_{x} \\ \hline & S_{z} & S_{y} \\ \hline & & S_{z} \\\end{array}\right]$}\rule[-18pt]{0pt}{40pt} \\ \hline
    $Y_{3a}$ & \scalebox{1}{$\begin{bmatrix}\begin{array}{c|c|c} O_{xy}^{y}, O_{zx}^{z}, \bm{O}^{x} & O_{xy}^{x}, O_{yz}^{z}, \bm{O}^{y} & O_{zx}^{x}, O_{yz}^{y}, O_{xx}^{z}, O_{yy}^{z} \\ \hline & O_{xy}^{y}, O_{zx}^{z}, \bm{O}^{x} & \bm{O}_{lm}^{\alpha} \\ \hline & & O_{xx}^{x}, O_{yy}^{x}, O_{xy}^{y}\end{array}\end{bmatrix}$} & \scalebox{1}{$\left[\begin{array}{c|c|c} S_{x} & S_{y} & - \\ \hline & S_{x} & - \\ \hline & & - \\\end{array}\right]$}\rule[-18pt]{0pt}{40pt} \\ \hline
    $Y_{3b}$ & \scalebox{1}{$\begin{bmatrix}\begin{array}{c|c|c} O_{xy}^{x}, O_{yz}^{z}, \bm{O}^{y} & O_{xy}^{y}, O_{zx}^{z}, \bm{O}^{x} & \bm{O}_{lm}^{\alpha} \\ \hline & O_{xy}^{x}, O_{yz}^{z}, \bm{O}^{y} & O_{zx}^{x}, O_{yz}^{y}, O_{xx}^{z}, O_{yy}^{z} \\ \hline & & O_{xy}^{x}, O_{xx}^{y}, O_{yy}^{y}\end{array}\end{bmatrix}$} & \scalebox{1}{$\left[\begin{array}{c|c|c} S_{y} & S_{x} & - \\ \hline & S_{y} & - \\ \hline & & - \\\end{array}\right]$}\rule[-18pt]{0pt}{40pt} \\
\hline
  \end{tabular}
  \label{tab:MO spin Ohmic tesseral}
\end{table*}
  \begin{table*}
  \caption{\raggedright
    Symmetry-allowed MO Ohmic conductivity tensors with activations of rank-4 and -5 \emph{t}-multipoles (\emph{t}-MP).
    It is noted that a corresponding spin conductivity tensor does not exist.
  }  
  \centering
  \begin{tabular}{c|c|c|c}
    \emph{t}-MP & MO conductivity & \emph{t}-MP & MO conductivity \\ \hline \hline
    $X_{40}$ & \scalebox{0.9}{$\begin{bmatrix}\begin{array}{c|c|c} \bm{O}_{lm}^{\alpha} & O_{zx}^{x}, O_{yz}^{y}, O_{xx}^{z}, O_{yy}^{z} & O_{xy}^{x}, O_{yz}^{z}, \bm{O}^{y} \\ \hline & \bm{O}_{lm}^{\alpha} & O_{xy}^{y}, O_{zx}^{z}, \bm{O}^{x} \\ \hline & & O_{yz}^{x}, O_{zx}^{y}\end{array}\end{bmatrix}$} & $Y_{50}$ & \scalebox{0.9}{$\begin{bmatrix}\begin{array}{c|c|c} O_{zx}^{x}, O_{yz}^{y}, \bm{O}^{z} & \bm{O}_{lm}^{\alpha} & O_{xy}^{y}, O_{zx}^{z}, \bm{O}^{x} \\ \hline & O_{zx}^{x}, O_{yz}^{y}, \bm{O}^{z} & O_{xy}^{x}, O_{yz}^{z}, \bm{O}^{y} \\ \hline & & O_{zx}^{x}, O_{yz}^{y}, \bm{O}^{z}\end{array}\end{bmatrix}$}\rule[-15pt]{0pt}{35pt}\\ \hline
    $X_{4a}$ & \scalebox{0.9}{$\begin{bmatrix}\begin{array}{c|c|c} O_{yy}^{x}, O_{zz}^{x}, O_{xy}^{y}, O_{zx}^{z} & O_{xy}^{x}, O_{yz}^{z}, \bm{O}^{y} & O_{zx}^{x}, O_{yz}^{y}, O_{xx}^{z}, O_{yy}^{z} \\ \hline & O_{xy}^{y}, O_{zx}^{z}, \bm{O}^{x} & \bm{O}_{lm}^{\alpha} \\ \hline & & O_{xx}^{x}, O_{yy}^{x}, O_{xy}^{y}\end{array}\end{bmatrix}$} & $Y_{5a}$ & \scalebox{0.9}{$\begin{bmatrix}\begin{array}{c|c|c} O_{xy}^{y}, O_{zx}^{z}, \bm{O}^{x} & O_{xy}^{x}, O_{yz}^{z}, \bm{O}^{y} & O_{zx}^{x}, O_{yz}^{y}, O_{xx}^{z}, O_{yy}^{z} \\ \hline & O_{xy}^{y}, O_{zx}^{z}, \bm{O}^{x} & \bm{O}_{lm}^{\alpha} \\ \hline & & O_{xx}^{x}, O_{yy}^{x}, O_{xy}^{y}\end{array}\end{bmatrix}$}\rule[-15pt]{0pt}{35pt}\\
\hline
  \end{tabular}
  \label{tab:MO Ohmic tesseral}
\end{table*}
  \begin{table*}
  \caption{\raggedright
    Symmetry-allowed MO Hall conductivity tensors with activations of rank-3 and -4 \emph{t}-multipoles (\emph{t}-MP).
    It is noted that a corresponding spin conductivity tensor does not exist.
  }
  \centering
  \begin{tabular}{c|c|c|c}
    \emph{t}-MP & MO conductivity & \emph{t}-MP & MO conductivity \\ \hline \hline
    $Y^{\alpha}_{z}$ & \scalebox{1}{$\begin{bmatrix}\begin{array}{c|c|c} - & O_{yz}^{x}, O_{zx}^{y} & O_{xy}^{y}, O_{zx}^{z}, \bm{O}^{x} \\ \hline & - & O_{xy}^{x}, O_{yz}^{z}, \bm{O}^{y} \\ \hline & & - \end{array}\end{bmatrix}$} & $X_{40}$ & \scalebox{1}{$\begin{bmatrix}\begin{array}{c|c|c} - & O_{zx}^{x}, O_{yz}^{y}, \bm{O}^{z} & O_{xy}^{x}, O_{yz}^{z}, \bm{O}^{y} \\ \hline & - & O_{xy}^{y}, O_{zx}^{z}, \bm{O}^{x} \\ \hline & & - \end{array}\end{bmatrix}$}\rule[-18pt]{0pt}{40pt}\\ \hline
    $Y_{3a}$ & \scalebox{1}{$\begin{bmatrix}\begin{array}{c|c|c} - & O_{xy}^{x}, O_{xx}^{y}, O_{yy}^{y} & O_{zx}^{x}, O_{yz}^{y}, O_{xx}^{z}, O_{yy}^{z} \\ \hline & - & \bm{O}_{lm}^{\alpha} \\ \hline & & - \end{array}\end{bmatrix}$} & $X_{4a}$ & \scalebox{1}{$\begin{bmatrix}\begin{array}{c|c|c} - & O_{xy}^{x}, O_{xx}^{y}, O_{yy}^{y} & O_{zx}^{x}, O_{yz}^{y}, O_{xx}^{z}, O_{yy}^{z} \\ \hline & - & \bm{O}_{lm}^{\alpha} \\ \hline & & - \end{array}\end{bmatrix}$}\rule[-18pt]{0pt}{40pt}\\ \hline
    $Y_{3b}$ & \scalebox{1}{$\begin{bmatrix}\begin{array}{c|c|c} - & O_{xx}^{x}, O_{yy}^{x}, O_{xy}^{y} & \bm{O}_{lm}^{\alpha} \\ \hline & - & O_{zx}^{x}, O_{yz}^{y}, O_{xx}^{z}, O_{yy}^{z} \\ \hline & & - \end{array}\end{bmatrix}$} & $X_{4b}$ & \scalebox{1}{$\begin{bmatrix}\begin{array}{c|c|c} - & O_{xx}^{x}, O_{yy}^{x}, O_{xy}^{y} & \bm{O}_{lm}^{\alpha} \\ \hline & - & O_{zx}^{x}, O_{yz}^{y}, O_{xx}^{z}, O_{yy}^{z} \\ \hline & & - \end{array}\end{bmatrix}$}\rule[-18pt]{0pt}{40pt}\\
\hline
  \end{tabular}
  \label{tab:MO Hall tesseral}
\end{table*}
  
  \section{\label{app:Hopping Hamiltonian}Hopping Hamiltonian}
  Here, the hopping matrices with basis functions $\{p_{x}, p_{y}, p_{z}, f_{x}^{\beta}, f_{y}^{\beta}, f_{z}^{\beta}\}$ are given below for each parameter $\{(ppt), (ppt), (pft), (pft), (fft), (fft), (fft), (fft)\}$, presented in the following form:
  \begin{equation}
    %
,
  \end{equation}
  where $t \in \{\sigma, \pi, \delta, \phi\}$.
  
  \subsection{First-nearest-neighbor hopping Hamiltonian}
  The abbreviation as $\mathrm{c}{\left(k_{i} \right)} = \cos k_{i}$, $\mathrm{s}{\left(k_{i} \right)} = \sin k_{i}$ for $i=x, y, z$ is used for notational simplicity.
  \begin{equation}
  H_{pp}^{\sigma} = 
  2 (pp\sigma)
  %
\end{equation}
  
  \subsection{Second-nearest-neighbor hopping Hamiltonian}
  \begin{equation}
  H_{pp}^{\sigma} = 
  2 (pp\sigma)
  %
\end{equation}
\end{widetext}

\bibliography{main}

\end{document}